\documentclass[a4paper,11pt]{article}
\usepackage{jheppub}
\usepackage[T1]{fontenc}
\usepackage[caption=false,font=normalsize,labelfont=sf,textfont=sf]{subfig}
\usepackage{tcolorbox}
\usepackage{tikz}
\usetikzlibrary{arrows}

\title{Geodesic equation in non-commutative gauge theory of gravity}

\author[a,b]{Abdellah Touati,\note{Corresponding author.}}
\author[b]{Slimane Zaim}

\affiliation[a]{Laboratoire de Physique des Rayonnements et de leurs Int\'{e}ractions avec la Mati\`{e}re\\
D\'{e}partement de Physique, Facult\'{e} des Sciences de la Mati\`{e}re\\
Universit\'{e} de Batna-1, Batna 05000, Algeria}
\affiliation[b]{D\'{e}partement de Physique, Facult\'{e} des Sciences de la Mati\`{e}re\\
Universit\'{e} de Batna-1, Batna 05000, Algeria}

\emailAdd{touati.abph@gmail.com}
\emailAdd{zaim69slimane@yahoo.com}

\abstract{In this work, we construct a non-commutative (NC) gauge theory of gravity for any metric with spherical symmetries, where we use a non-diagonal tetrad field. The deformed gauge potentials (tetrad fields) and the components of deformed metric are computed to the second order in the NC parameter $\Theta^{\mu\nu}$, as the application to the Schwarzschild black hole we show that the NC geometry removes the singularity at the origin of the black hole, and increase the event horizon. The non-commutativity correction to the effective potential of the Schwarzschild metric is also computed and we show how this geometry affects the stability condition which it found the NC parameter plays the same role as the mass that can be used to explain the dark matter and we show that the NC Schwarzschild space-time has new stable circular orbits appear near the event horizon that is not allowed by Schwarzschild space-time. The geodesic equations in the NC space and the corrections to the periastron advance in terms of $\Theta$ are obtained. We have also specified the problem of Mercury's perihelion and used the experimental data to estimate the NC parameter $\Theta$, then we show that $\Theta$ of the order $10^{-25}s.kg^{-1}$ gives observable corrections to the movement at a large scale. We show that the NC propriety of the spacetime appears at the High Energy.}

\keywords{Non-commutative geometry, Gauge gravity, Schwarzschild space-time, Geodesic equation.}

\begin{document}
\maketitle
\flushbottom

\section{Introduction}
In general relativity, the study of the geodesic motion of test particles in curved spacetime is the first probe towards an understanding of the physics and geometry of gravitational objects. So that she was able to solve the problem of mercury perihelion and is considered one of the triumphs of the general relativity theory. Also, the classification of stable and unstable orbits is done by drawing the effective potentials and checking their behavior at different points. We find many references that studied in detail the geodesy movement of black holes \cite{Chandr1,gibbons1,islam1,jaklit,stuchli,kran1,kran2,cruz,hackmann1,kagraman,hackmann2,hackmann3,levin,grunau,belbrun,barack,pugliese1,pugliese2,virb1,virb2,pradhan,olivares,villanueva,pugl1,pugl2}.. However, our interest is in studying the geodesic motion in non-commutative space-time, by imposing further commutation relations between position coordinates themselves. This non-commutativity leads to the modification of Heisenberg uncertainty relations in such a way that prevents one from measuring positions to better accuracy than the Planck length. Noncommutativity is mainly motivated by string theory, being a limit in the presence of a background field \cite{hawking1,hawking2,hod,vaz,harms,gour,zaim1,mebarki1}. This idea results in the concept of quantum gravity since quantifying space-time leads to quantifying gravity, these quantum gravity effects can be neglected in low energy limits, but in the strong gravitational field of a black hole one has to consider these effects.
In the non-commutative space-time the coordinate and momentum operators satisfy the following commutation relations:
\begin{equation}
	[x^{\mu},x^{\nu}]=i\Theta^{\mu\nu}
\end{equation}
where $\Theta^{\mu\nu}$ is an anti-symmetric real matrix which determines the fundamental cell discretization of space-time much in the same way as the Planck constant $\hbar$ discretizes the phase space and $x^{\mu}$ are the coordinate operator on a non-commutative space-time is defined by the following transformations:
\begin{equation}
	\hat{x}^{\mu}=x^{\mu}-\Theta^{\mu\nu}p_{\nu}
\end{equation}
In the non-commutative theory, the ordinary product has changed to the star product (Moyal product) "$*$" between two arbitrary functions $f(x)$ and $g(x)$ defined over this space-time:
\begin{equation}
	(f*g)(x)=f(x)e^{\frac{i}{2}\Theta^{\mu\nu}\overleftarrow{\partial_{\mu}}\overrightarrow{\partial_{\nu}}}g(x)\label{eqt2.25}
\end{equation}
Recently there has been interest in studies that investigated the modifications introduced by a non-commutativity on the geodesic structure in black hole \cite{nozari1,nozari2,kuniyal,mirza,rome1,iftikhar,ulh1,larranaga,nozari,nicolini1,bhar,rahaman}.
Here one uses a gauge gravity theory in non-commutative space-time with star products and Seiberg-Witten maps \cite{seiberg1}. The non-commutative gauge gravity is a theory of general relativity in curved space-time with preservation of non-commutative space-time and based partly on implementing symmetries on flat non-commutative space-time. In gauge gravity theory the action transforms under ordinary Lorentz transformation of the ordinary fields, since these ordinary transformations, via the Seiberg-Witten map, induce the non-commutative canonical transformations of non-commutative fields under which the non-commutative action is invariant \cite{chai1,haranas1,haranas2}. Our work will be in this context, with the aim of writing the geodesic equation that arises from the metric tensor which has been corrected using star product between tetrads fields and Seiberg-Witten maps. In addition, we hope to obtain corrections in terms of the non-commutative for each of the effective potential and the deviation angle per revolution and we also discussed the issue of the stability of circular orbits in a non-commutative Schwarzschild geometry. The article is organized as follows. In Sect. 2 we present the non-commutative corrections to the metric field using star product between tetrads fields and Seiberg-Witten maps. In Sect. 3 we calculate the non-commutative geodesic equation and we also obtained non-commutative effective potentials up to the second-order from the non-commutativity parameter and the condition for the stability of the circular orbitals of the particles in non-commutative Schwarzschild space-time has also been determined. We then calculated the non-commutative adjustment to the value of the perihelion rotation and gave an estimate for the non-commutative parameter. In the last section, we present our concluding remarks.


\section{Non-commutative gauge gravity for spherical symmetric metric}

Using the tetrad and spin connection formalism in the gauge theory of gravity is unavoiable because of the requirement to describe the spinor fields in this theory. We denote the tetrad fields by $e^{a}_{\mu},a=0,1,2,3$ and the spin connection $\omega^{ab}_{\mu}(x)=-\omega^{ba}_{\mu}(x)$ [ab]=[01],[02],[03],[12],[13],[23]. Then the Ricci scalar is given by:
\begin{equation}
	R=e_{a}^{\mu}e_{b}^{\nu}R^{ab}_{\mu\nu}
\end{equation}
Where $e_{a}^{\mu}$ denotes the inverse of $e^{a}_{\mu}$ which satisfying the usual properties:
\begin{equation}
	e^{a}_{\mu}e_{b}^{\mu}=\delta^{a}_{b},\quad e^{a}_{\mu}e_{a}^{\nu}=\delta^{\nu}_{\mu}
\end{equation}
And $R^{ab}_{\mu\nu}$ denote the curvature tensor:
\begin{equation}
	R^{ab}_{\mu\nu}=\partial_{\mu}\omega^{ab}_{\nu}-\partial_{\nu}\omega^{ab}_{\mu}+\left(\omega^{ac}_{\mu}\omega^{db}_{\nu}-\omega^{ac}_{\nu}\omega^{db}_{\mu}\right)\eta_{cd}\label{eqt2.3}
\end{equation}	
So then the action of the pur gravity in the gauge theory, are define  read:
\begin{equation}
	S_{g}=\frac{1}{16\pi G}\int d^{4}xe R=\frac{1}{16\pi G}\int d^{4}xe e_{a}^{\mu}e_{b}^{\nu}R^{ab}_{\mu\nu}\label{eqt2.1}
\end{equation}
where $e=det(e^{a}_{\mu})$.
Using the variational principle $\delta S=0$ for the action in \eqref{eqt2.1} respect to $e^{a}_{\mu}$, we can get the field equation for the gravitational potentials $e^a_\mu$ in the vacuum:
\begin{equation}
	R^a_{\mu}=0\label{eqt2.5}
\end{equation}
where $	R^a_{\mu}=R^{ab}_{\mu\nu}e^{\nu}_b$ the Ricci tensor.

We consider the general solution of the equation of the gravitational field \eqref{eqt2.5} in the case of static and spherical symmetry, the following metric:
\begin{equation}
	ds^{2}=-A(r)dt^{2}+B(r)dr^{2}+r^{2}(d\theta^{2}+sin^{2}\theta d\phi^{2})\label{eqt2.6}
\end{equation}
where the $A(r)$ and $B(r)$ are functions related only to the radius $r$. The tetrad formulation of General Relativity allows to write the tetrad components with this relation:
\begin{equation}
	g_{\mu\nu}=e^{a}_{\mu}e_{a\nu}\label{eqt2.7}
\end{equation}

With the choose a particular form of non diagonal tetrad fields satisfies the relation \eqref{eqt2.7} as follows:
\begin{equation}
	e^{a}_{\mu}=\left[\begin{array}{cccc}
		A(r) & 0 & 0 & 0 \\
		0 & B(r)sin\theta cos\phi & r cos\theta cos\phi & -r sin\theta sin\phi \\
		0 & B(r)sin\theta sin\phi & r cos\theta sin\phi & r sin\theta cos\phi \\
		0 & B(r)cos\theta & -r sin\theta & 0
	\end{array} \right]	\label{eqt2.8}
\end{equation}
We note that, this particular form of tetrad field can be used for a stationary observer at spatial infinity \cite{ulh1}.

The non-zero component of the spin connection for this tetrad field:
\begin{align}
	\omega^{01}_{\mu}&=\left(\frac{A'(r)}{B(r)}sin\theta cos\phi,0,0,0\right),\omega^{02}_{\mu}=\left(\frac{A'(r)}{B(r)}sin\theta sin\phi,0,0,0\right),\\\omega^{03}_{\mu}&=\left(\frac{A'(r)}{B(r)}cos\theta ,0,0,0\right),\omega^{12}_{\mu}=\left(0,0,0,[1-\frac{1}{B(r)}]sin^{2}\theta\right),\\\omega^{13}_{\mu}&=\left(0,0,-[1-\frac{1}{B(r)}]cos\phi,[1-\frac{1}{B(r)}]sin\theta cos\theta sin\phi\right),\\\omega^{23}_{\mu}&=\left(0,0,-[1-\frac{1}{B(r)}]sin\phi,-[1-\frac{1}{B(r)}]sin\theta cos\theta cos\phi\right),	
\end{align}

Using the relations \eqref{eqt2.3}, the spin connection, and the tetrads fields, we find the nonzero component of the curvature tensor $R^{ab}_{\mu\nu}$, which we need to use further with the spin connection and tetrads fields, in the derivation of the expressions of the deformed tetrads fields:
\begin{align}
	R^{01}_{01}&=-\left[\frac{A''(r)}{B(r)}-\frac{A'(r)B'(r)}{B^{2}(r)}\right]sin\theta cos\phi,\quad R^{01}_{02}=-\frac{A'(r)}{B^{2}(r)}cos\theta cos\phi,\\
	R^{01}_{03}&=\frac{A'(r)}{B^{2}(r)}sin\theta sin\phi,\quad R^{02}_{01}=-\left[\frac{A''(r)}{B(r)}-\frac{A'(r)B'(r)}{B^{2}(r)}\right]sin\theta sin\phi ,\\R^{02}_{02}&=-\frac{A'(r)}{B^{2}(r)}cos\theta sin\phi,\quad
	R^{02}_{03}=-\frac{A'(r)}{B^{2}(r)}sin\theta cos\phi,\quad R^{03}_{02}=\frac{A'(r)}{B^{2}(r)}sin\theta ,\\R^{03}_{01}&=-\left[\frac{A''(r)}{B(r)}-\frac{A'(r)B'(r)}{B^{2}(r)}\right]cos\theta,\quad R^{12}_{23}=\left[1-\frac{1}{B^{2}(r)}\right]sin\theta cos\theta\\ R^{12}_{13}&=\frac{B'(r)}{B^{2}(r)}sin^{2}\theta,\quad R^{13}_{12}=-\frac{B'(r)}{B^{2}(r)}cos\phi,\quad R^{13}_{13}=\frac{B'(r)}{B^{2}(r)}sin\theta cos\theta sin\phi,\\
	R^{13}_{23}&=-\left[1-\frac{1}{B^{2}(r)}\right]sin^{2}\theta sin\phi,\quad R^{23}_{13}=-\frac{B'(r)}{B^{2}(r)}sin\theta cos\theta cos\phi,\\ R^{23}_{12}&=-\frac{B'(r)}{B^{2}(r)}sin\phi,\quad R^{23}_{23}=\left[1-\frac{1}{B^{2}(r)}\right]sin^{2}\theta cos\phi,
\end{align}
where $A'(r)$, $B'(r)$ and $A''(r)$ denote the derivatives of first and second-order with respect to the r-coordinate.

In the non-commutative space-time, for finding the deformed tetrad fields $\hat{e}^{a}_{\mu}(x,\Theta)$ we use the Seiberg-Witten map, which describes the tetrad fields as a development in the power of $\Theta$ up to the second-order \cite{cham1}:
\begin{equation}
	\hat{e}^{a}_{\mu}(x,\Theta)=e^{a}_{\mu}(x)-i\Theta^{\nu\rho}e^{a}_{\mu\nu\rho}(x)+\Theta^{\nu\rho}\Theta^{\lambda\tau}e^{a}_{\mu\nu\rho\lambda\tau}(x)+\mathcal{O}(\Theta^{3})\label{eqt2.26}
\end{equation}
where:
\begin{align}
	e^{a}_{\mu\nu\rho}&=\frac{1}{4}[\omega^{ac}_{\nu}\partial_{\rho}e^{d}_{\mu}+(\partial_{\rho}\omega^{ac}_{\mu}+R^{ac}_{\rho\mu})e^{d}_{\nu}]\eta_{cd}\\
	e^{a}_{\mu\nu\rho\lambda\tau}&=\frac{1}{32}\left[2\{R_{\tau\nu},R_{\mu\rho}\}^{ab}e^{c}_{\lambda}-\omega^{ab}_{\lambda}(D_{\rho}R_{\tau\nu}^{cd}+\partial_{\rho}R_{\tau\nu}^{cd})e^{m}_{\nu}\eta_{dm}\right.\notag\\
	&\left.-\{\omega_{\nu},(D_{\rho}R_{\tau\nu}+\partial_{\rho}R_{\tau\nu})\}^{ab}e^{c}_{\lambda}-\partial_{\tau}\{\omega_{\nu},(\partial_{\rho}\omega_{\mu}+R_{\rho\mu})\}^{ab}e^{c}_{\lambda}\right.\notag\\
	&\left.-\omega^{ab}_{\lambda}\left(\omega^{cd}_{\nu}\partial_{\rho}e^{m}_{\mu}+\left(\partial_{\rho}\omega_{\mu}^{cd}+R_{\rho\mu}^{cd}\right)e^{m}_{\nu}\right)\eta_{dm}+2\partial_{\nu}\omega_{\lambda}^{ab}\partial_{\rho}\partial_{\tau}e^{c}_{\lambda}\right.\notag\\
	&\left.-2\partial_{\rho}\left(\partial_{\tau}\omega_{\mu}^{ab}+R_{\tau\mu}^{ab}\right)\partial_{\nu}e^{c}_{\lambda}-\{\omega_{\nu},(\partial_{\rho}\omega_{\lambda}+R_{\rho\lambda})\}^{ab}\partial_{\tau}e^{c}_{\mu}\right.\notag\\
	&\left.-\left(\partial_{\tau}\omega_{\mu}+R_{\tau\mu}\right)\left(\omega^{cd}_{\nu}\partial_{\rho}e^{m}_{\lambda}+\left((\partial_{\rho}\omega_{\lambda}+R_{\rho\lambda})\right)e^{m}_{\nu}\right)\eta_{dm}\right]\eta_{cb}\label{eqt2.28}		
\end{align}
and
\begin{align}
	\{\alpha,\beta\}^{ab}=\left(\alpha^{ac}\beta^{db}+\beta^{ac}\alpha^{db}\right)\eta_{cd},\quad &[\alpha,\beta]^{ab}=\left(\alpha^{ac}\beta^{db}-\beta^{ac}\alpha^{db}\right)\eta_{cd}\\
	D_{\mu}R_{\rho\sigma}^{ab}=\partial_{\mu}R^{ab}_{\rho\sigma}+&\left(\omega_{\mu}^{ac}R^{db}_{\rho\sigma}+\omega_{\mu}^{bc}R^{da}_{\rho\sigma}\right)
\end{align}

The complex conjugate $\hat{e}^{a\dagger}_{\mu}(x,\Theta)$ of the deformed tetrad fields is obtien from the hermitian conjugate of the relation \eqref{eqt2.26}:
\begin{equation}
	\hat{e}^{a\dagger}_{\mu}(x,\Theta)=e^{a}_{\mu}(x)+i\Theta^{\nu\rho}e^{a}_{\mu\nu\rho}(x)+\Theta^{\nu\rho}\Theta^{\lambda\tau}e^{a}_{\mu\nu\rho\lambda\tau}(x)+\mathcal{O}(\Theta^{3})\label{eqt2.32}
\end{equation}

and the real deformed metric is given by the formula \cite{chai1}:
\begin{equation}
	\tilde{g}_{\mu\nu}(x,\Theta)=\frac{1}{2}\left[\hat{e}^{a}_{\mu}*\hat{e}^{b\dagger}_{\nu}+\hat{e}^{a}_{\nu}*\hat{e}^{b\dagger}_{\mu}\right]\eta_{ab}\label{eqt2.33}
\end{equation}

Using the Seiberg-Witten map \eqref{eqt2.28}, we can easily get the deformed tetrad fields $\hat{e}^{a}_{\mu}(x,\Theta)$ and he's hermitian conjugate $\hat{e}^{a\dagger}_{\mu}(x,\Theta)$ given by the relations \eqref{eqt2.26} and \eqref{eqt2.32}.
To simplify the calculations, we took only space-space non-commutativity, $\Theta_{0i}=0$ (due to the known problem with unitary), so we choose the following the metric for the non-commutativity parameter $\Theta^{\mu\nu}$:
\begin{equation}
	\Theta^{\mu\nu}=\left(\begin{matrix}
		0	& 0 & 0 & 0 \\
		0	& 0 & 0 & \Theta \\
		0	& 0 & 0 & 0 \\
		0	& -\Theta & 0 & 0
	\end{matrix}
	\right), \qquad \mu,\nu=0,1,2,3\label{eqt2.34}
\end{equation}
where $\Theta$ is a real positive constant.

The non-zero components of the non-commutative tetrad fields $\hat{e}^{a}_{\mu}$ are:

\begin{align}
	\hat{e}^{0}_{0}=&A(r)+\frac{\Theta^{2}}{32B^{4}(r)}\left\{-4B(r)(4rB'(r)A''(r)+A'(r)(2B'(r)+rB''(r)))\right.\notag\\
	&\left.+16rA'(r)B'^{2}(r)+B^{3}(r)(A'(r)B'(r)+4A''(r))+B^{2}(r)(-3A'(r)B'(r)\right.\notag\\
	&\left.+4(A''(r)+rA'''(r)))\right\} sin^{2}\theta+\mathcal{O}(\Theta^{3})\\
	\hat{e}^{1}_{1}=&B(r)sin\theta cos\phi+\frac{i\Theta}{4}B'(r)sin\theta sin\phi+\frac{\Theta^{2}}{64B^{3}(r)}\left\{8(2B'(r)-B(r)B''(r))sin^{2}\theta\right.\notag\\
	&\left.+B^{3}(r)B''(r)(3+cos2\theta)+B(r)(B'^{2}(r)-B(r)B''(r))(1+3cos2\theta)\right\}sin\theta cos\phi\notag\\
	&+\mathcal{O}(\Theta^{3})\\
	\hat{e}^{2}_{1}=&B(r)sin\theta sin\phi-\frac{i\Theta}{4}B'(r)sin\theta cos\phi+\frac{\Theta^{2}}{64B^{3}(r)}\left\{8(2B'(r)-B(r)B''(r))sin^{2}\theta\right.\notag\\
	&\left.+B^{3}(r)B''(r)(3+cos2\theta)+B(r)(B'^{2}(r)-B(r)B''(r))(1+3cos2\theta)\right\}sin\theta sin\phi\notag\\
	&+\mathcal{O}(\Theta^{3})\\
	\hat{e}^{3}_{1}=&\frac{\Theta^{2}sin^{2}\theta}{32B^{3}(r)}\left\{(8-3B(r))B'^{2}(r)-B(r)B''(r)(4+(-3+B(r))B(r)) \right\}cos\theta\notag\\
	&+B(r)cos\theta+\mathcal{O}(\Theta^{3})\\
	\hat{e}^{1}_{2}=&rcos\theta cos\phi-\frac{i\Theta}{4}\left[B(r)-1\right]cos\theta sin\phi+\frac{\Theta^{2}}{32B^{4}(r)}\left\{B^{4}(r)B'(r)(-3+cos2\theta)\right.\notag\\
	&\left.+sin^{2}\theta\left[16rB'^{2}(r)-B^{2}(r)(B'(r)-4rB''(r))-4B(r)(2B'(r)+2rB'^{2}(r)+rB''(r))\right]\right.\notag \\
	&\left.-\frac{1}{2}B^{3}(r)B'(r)(-9+5cos2\theta)\right\}cos\theta cos\phi+\mathcal{O}(\Theta^{3})
\end{align}
\begin{align}
	\hat{e}^{2}_{2}=&rcos\theta sin\phi+\frac{i\Theta}{4}\left[B(r)-1\right]cos\theta cos\phi+\frac{\Theta^{2}}{32B^{4}(r)}\left\{B^{4}(r)B'(r)(-3+cos(2\theta))\right.\notag\\
	&\left.+sin^{2}\theta\left[16rB'^{2}(r)-B^{2}(r)(B'(r)-4rB''(r))-4B(r)(2B'(r)+2rB'^{2}(r)+rB''(r))\right]\right.\notag \\
	&\left.-\frac{1}{2}B^{3}(r)B'(r)(-9+5cos(2\theta))\right\}cos\theta sin\phi+\mathcal{O}(\Theta^{3}) \\
	\hat{e}^{3}_{2}=&-rsin\theta +\frac{\Theta^{2}sin\theta}{64B^{4}(r)}\left\{sin^{2}\theta\left[4B(r)B'(r)(4+B^{3}(r))-32rB'^{2}(r)+8rB(r)B''(r) \right] \right.\notag\\
	&\left.B^{2}(r)B'(r)(5-B(r)+(-1+5B(r))cos(2\theta))+8rB(r)(B(r)B''(r)-2B'^{2}(r))cos^{2}\theta\right\}\notag \\
	&+\mathcal{O}(\Theta^{3}) \\
	\hat{e}^{1}_{3}=&-rsin\theta sin\phi-\frac{i\Theta}{4}\left[\left(B(r)-1\right)cos^{2}\theta-((1-\frac{1}{B(r)})+2\frac{B'(r)}{B^{2}(r)}r)sin^{2}\theta\right]sin\theta cos\phi\notag \\
	&+\frac{\Theta^{2}}{32B^{4}(r)}\left\{[+3B^{2}(r)B'(r)+36rB'^{2}(r)+8rB^{2}(r)B''(r)-B(r)(7B'(r)+16rB'^{2}(r)\right.\notag\\
	&\left.+12rB''(r))]sin^{2}\theta+2B^{3}(r)B'(r)-2B^{4}(r)B'(r)cos^{2}\theta\right\}(-sin\theta sin\phi)+\mathcal{O}(\Theta^{3})\\
	\hat{e}^{2}_{3}=&rsin\theta cos\phi+\frac{i\Theta}{4}\left[\left(B(r)-1\right)cos^{2}\theta-((1-\frac{1}{B(r)})+2\frac{B'(r)}{B^{2}(r)}r)sin^{2}\theta\right](-sin\theta sin\phi)\notag \\
	&+\frac{\Theta^{2}}{32B^{4}(r)}\left\{[+3B^{2}(r)B'(r)+36rB'^{2}(r)+8rB^{2}(r)B''(r)-B(r)(7B'(r)+16rB'^{2}(r)\right.\notag\\
	&\left.+12rB''(r))]sin^{2}\theta+2B^{3}(r)B'(r)-2B^{4}(r)B'(r)cos^{2}\theta\right\}(sin\theta cos\phi)+\mathcal{O}(\Theta^{3})\\
	\hat{e}^{3}_{3}=&\frac{i\Theta}{4B(r)^2}\left[(-B(r)+B(r)^3+2rB'(r))\right]sin^2\theta cos\theta
\end{align}

Then, using the definition \eqref{eqt2.33}, to obtain the non-zero components of the non-commutative metric $\tilde{g}_{\mu\nu}$ up to the second-order of $\Theta$, and we intend to analyze a geodesic movement over a plane $\theta=\frac{\pi}{2}$, so the new metric will assume a simpler diagonal form:
\begin{align}
	\tilde{g}_{00}=&-A^{2}(r)-\frac{A(r)\Theta^{2}}{16B^{4}(r)}\left\{-4B(r)(4rB'(r)A''(r)+A'(r)(2B'(r)+rB''(r)))\right.\notag\\
	&\left.+16rA'(r)B'^{2}(r)+B^{3}(r)(A'(r)B'(r)+4A''(r))+B^{2}(r)(-3A'(r)B'(r)\right.\notag\\
	&\left.+4(A''(r)+rA'''(r)))\right\}+\mathcal{O}(\Theta^{3})\label{eqt2.44}\\
	\tilde{g}_{11}=&B^{2}(r)+\frac{\Theta^{2}}{16B^{2}(r)}\left\{B'^{2}(r)\left(8+B(r)(-1+9B(r))\right)+B''(r)B(r)(-4+B(r)\right.\notag\\
	&\left.+9B^{2}(r))\right\}+\mathcal{O}(\Theta^{3})	\label{eqt2.45}\\
	\tilde{g}_{22}=&r^{2}+\frac{\Theta^{2}}{16B^4(r)}\left\{B'(r)\left(-B(r)(1+B(r))+(8+B(r)(-5+2B(r)+16rB'(r)))\right)\right.\notag\\
	&\left.-4rB(r)B'(r)\right\}r+\mathcal{O}(\Theta^{3})\label{eqt2.46}\\
	\tilde{g}_{33}=&r^{2}+\frac{\Theta^{2}}{16B^4(r)}\left\{9B^{4}(r)-2B^{3}(r)(3-rB'(r))+40r^{2}B'^{2}(r)-rB(r)(B'(r)(11+32rB'(r))\notag\right.\\
	&\left.+12rB''(r))+B^{2}(r)(1+27rB'(r)+16r^{2}B''(r))\right\}+\mathcal{O}(\Theta^{3})\label{eqt2.47}
\end{align}
We can clearly see that, if $\Theta\rightarrow 0$, we obtain the commutative metric \eqref{eqt2.6}.


\section{Geodesic equation in the non-commutative Schwarzschild spacetime}

The structure of spacetime in the non-commutative case it has given by the new line element:	
\begin{align}
	ds^{2}&=\tilde{g}_{00}(r,\Theta)c^{2}dt^{2}+\tilde{g}_{11}(r,\Theta)dr^{2}+\tilde{g}_{22}(r,\Theta)d\theta^{2}+ \tilde{g}_{33}(r,\Theta)d\phi^{2}\label{eqt3.1}
\end{align}		
Now, if we insert the Schwarzschild potential $A(r)=B^{-1}(r)=\left(1-\frac{2 m}{r}\right)^{\frac{1}{2}}$ into (\ref{eqt2.44}-\ref{eqt2.45}-\ref{eqt2.46}-\ref{eqt2.47}), then we obtien the deformed Schwarzschild metric with a correction up to the second order in $\Theta$:
\begin{align}
	-\tilde{g}_{00}=&\left(1-\frac{2 m}{r}\right)+\left\{\frac{m\left(88m^2+mr\left(-77+15\sqrt{1-\frac{2m}{r}}\right)-8r^2\left(-2+\sqrt{1-\frac{2m}{r}}\right)\right)}{16 r^4(-2m+r)}\right\}\Theta^{2}+\mathcal{O}(\Theta^{3})\label{eqt2.48}\\
	\tilde{g}_{11}=&\left(1-\frac{2 m}{r}\right)^{-1}+\left\{\frac{m\left(12m^2+mr\left(-14+\sqrt{1-\frac{2m}{r}}\right)-r^2\left(5+\sqrt{1-\frac{2m}{r}}\right)\right)}{8r^2(-2m+r)^3}\right\}\Theta^{2}+\mathcal{O}(\Theta^{3})\label{eqt2.49}\\
	\tilde{g}_{22}=&r^{2}+\left\{\frac{m\left(m\left(10-6\sqrt{1-\frac{2m}{r}}\right)-\frac{8m^2}{r}+r\left(-3+5\sqrt{1-\frac{2m}{r}}\right)\right)}{16(-2m+r)^2}\right\}\Theta^{2}+\mathcal{O}(\Theta^{3})\label{eqt2.50}\\
	\tilde{g}_{33}=&r^{2}+\left\{\frac{5}{8}-\frac{3}{8}\sqrt{1-\frac{2m}{r}}+\frac{m\left(-17+\frac{5}{\sqrt{1-\frac{2m}{r}}}\right)}{16r}+\frac{m^2\sqrt{1-\frac{2m}{r}}}{(-2m+r)^2}\right\}\Theta^{2}+\mathcal{O}(\Theta^{3})\label{eqt2.51}
\end{align}
From these expressions, all the non-zero components of the metric acquire a singularity in the NC correction term at $r=2m$, also in $\tilde{g}_{00}$ component, that's what we don't see in Ref. \cite{chai1}, this difference was the result of using a general form of the tetrad filed.

The corresponding event horizon in the non-commutative Schwarzschild black hole can be obtain by solving $\tilde{g}_{tt}=0$.
\begin{equation}
	r_{H}^{NC}=r_{H}^{C}+\frac{1}{32}\sqrt{8-2r_{H}^{C}}\Theta-(\frac{13r_{H}^{C}-48}{256r_{H}^{C}})\Theta^2
\end{equation}
where $r_{H}^{C}=2m$ is the event horizon of the Schwarzschild black hole in the commutative spacetime when $\Theta=0$.

\begin{figure}[h]
	\centering
	\includegraphics[width=0.6\textwidth]{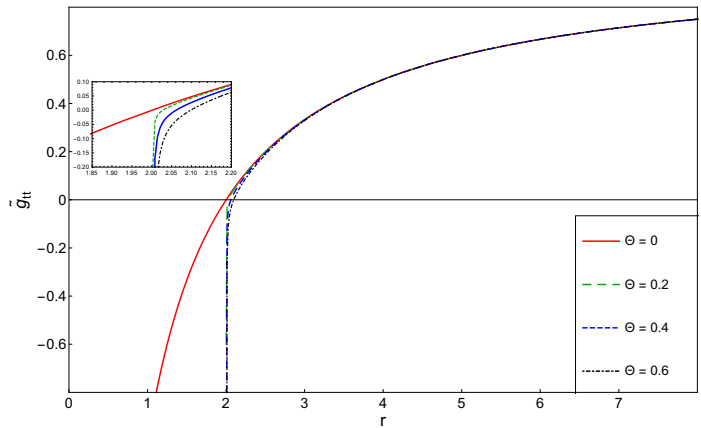}
	\caption{Behaviors of $\tilde{g}_{tt}$ for a stationary observer at spatial infinity in the non-commutative spacetime with a given $\Theta$.}	\label{fig1.1}
\end{figure}

As we see in Fig.\ref{fig1.1}, the event horizon in the NC spacetime is bigger than in the commutative case $r_{H}^{NC}>r_{H}^{C}$, and new behaviors of $\tilde{g}_{\mu\nu}$ are showing in this theory, where the singularity of the Schwarzschild solution at $r=0$ is now shifted by the noncommutativity of space to the finite radius $r=2m$. Then the
interior region ($r<r_{H}$) became banned from any observer, so the NC structure of the space-time gives a non-singular black hole. Which is a logical result because we are using a general form of the tetrad field for a stationary observer at the spatial infinity \eqref{eqt2.8}. This result is not available in the diagonal form of tetrad in the NC gauge theory as in the literature \cite{chai1,chai2,mukhe1} or in the theory of non-singularity black hole like the noncommutativity eliminates point-like gravitational source \cite{nozari,larranaga,nicolini1,bhar,rahaman}, Hayward black hole \cite{hayw,hayw1,hayw2,hayw3}. We can see this results in the theory of quantum-corrected black hole \cite{qcbh1,qcbh2,nozari3}, but just in the particular case where $a=r_{H}$, and $a$ in this theory represent a minimal distance and it is expected of the order of the Planck length, $l_{p}$, when the singularity of the black hole in this theory is shifted to $r=a=r_{min}\sim l_{p}$, so is not a natural result because we need to fixed the parameter $a$ for a particular value to observe the same result as in Fig \ref{fig1.1}. Contrary to our results, which is emerged naturally from the quantum structure of the space-time itself, when we impose the NC property of the geometry to the space-time, without the need to impose a particular value to the NC parameter $\Theta$.
Then we conclude that the NC geometry removes the singularity at the origin of the black hole and increases the radius of the event horizon.

The corresponding Lagrangian can be written according to the non-commutative spacetime structure described by \eqref{eqt3.1}, as follows.
\begin{align}
	2L=&\tilde{g}_{tt}(r,\Theta)c^{2}\dot{t}^{2}+\tilde{g}_{rr}(r,\Theta)\dot{r}^{2}+\tilde{g}_{\phi\phi}(r,\Theta)\dot{\phi}^{2}\label{eqt3.49}
\end{align}
where the dots represent the derivative with respect to the affine parameter $\tau$, along the geodesic.

Use the Euler-Lagrange equation:
\begin{equation}
	\frac{d}{ds}\left(\frac{\partial L}{\partial \dot{x}^{\mu}}\right)-\frac{\partial L}{\partial x^{\mu}}=0
\end{equation}

Since $L$ is independent of $t$ and $\phi$ so we have tow conserved quantities:
\begin{align}
	E_{0}&=p_{t}=c^{2}\tilde{g}_{tt}(r,\Theta)\dot{t}\Rightarrow \dot{t}=\frac{E_{0}}{c^{2}\tilde{g}_{tt}(r,\Theta)}\label{eqt3.59}\\
	l&=p_{\phi}=\tilde{g}_{\phi\phi}(r,\Theta)\dot{\phi}\Rightarrow \dot{\phi}=\frac{l}{\tilde{g}_{\phi\phi}(r,\Theta)}\label{eqt3.60}
\end{align}
Use the invariant\footnote{where $U^{\mu}=c^{-1}\frac{dx^{\mu}}{d\tau}$ denote the 4-velocity.} of $\tilde{g}_{\mu\nu}U^{\mu}U^{\nu}\equiv -h$, together with the relations \eqref{eqt3.59} and \eqref{eqt3.60} and with some rearrange we can get the explicit equation for $\dot{r}^{2}$:
\begin{equation}
	\dot{r}^{2}=-\frac{E_{0}^{2}}{c^{2}\tilde{g}_{tt}(r,\Theta)\tilde{g}_{rr}(r,\Theta)}-\frac{1}{\tilde{g}_{rr}(r,\Theta)}\left(\frac{l^{2}}{\tilde{g}_{\phi\phi}(r,\Theta)}+hc^{2}\right)\label{eqt3.11}
\end{equation}
where we shall consider $h=m_{0}^{2}$ for massive particles.

Substituting \eqref{eqt2.48}, \eqref{eqt2.49} and \eqref{eqt2.51} into \eqref{eqt3.11}, and with the developed up to $\mathcal{O}(\Theta^{3})$ respect to $\Theta$, then the equation \eqref{eqt3.11} can be written as:
\begin{equation}
	\dot{r}^{2}+V_{eff}(r,\Theta)=0 \label{eqt 3.68}
\end{equation}
where:
\small
\begin{align}
	V_{eff}(r,\Theta)=&\left(1-\frac{2 m}{r}\right)\left(\frac{l^{2}}{r^{2}}+hc^{2}\right)-E^2+\Theta^{2}\left\{-\frac{l^{2}}{r^{4}}\left(\frac{5}{8}-\frac{3}{8}\sqrt{1-\frac{2m}{r}}+\frac{m\left(-17+\frac{5}{\sqrt{1-\frac{2m}{r}}}\right)}{16r}\right.\right.\notag\\
	&\left.\left.+\frac{m^2\sqrt{1-\frac{2m}{r}}}{(-2m+r)^2}\right)+E^{2}\left(\frac{m(64m^2+m(-49+13\sqrt{1-\frac{2 m}{r}})r+2(13-3\sqrt{1-\frac{2 m}{r}})r^2)}{16r^5(1-\frac{2m}{r})^2}\right)\right.\notag\\
	&\left.+\left(\frac{l^{2}}{r^{2}}+hc^{2}\right)\left(\frac{m(12m^2+m(-14+\sqrt{1-\frac{2 m}{r}})r-(5+\sqrt{1-\frac{2 m}{r}})) r^2}{8r^5(1-\frac{2m}{r})}\right)\right\}+\mathcal{O}(\Theta^{4})\notag\\ \label{eqt3.73}
\end{align}
\normalsize
It is clear, for $\Theta\rightarrow 0$, we obtien the commutative effective potential for the Schwarzschild metric:
\begin{equation}
	V_{eff}(r,\Theta=0)=\left(1-\frac{2 m}{r}\right)\left(\frac{l^{2}}{r^{2}}+hc^{2}\right)-E^2
\end{equation}

\begin{figure}[h]
	
	\begin{center}
	\includegraphics[width=0.5\textwidth]{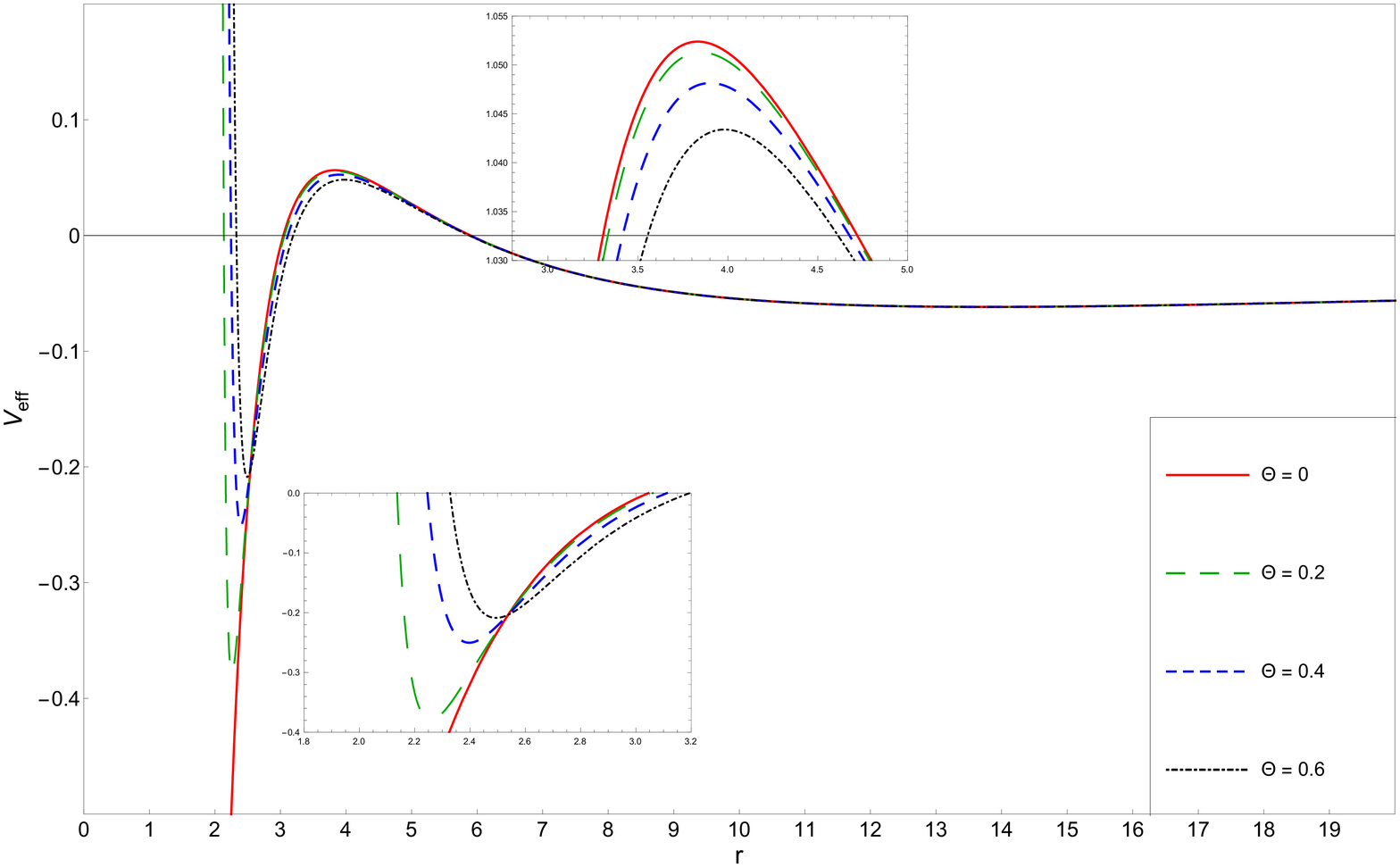}\hfill
	\includegraphics[width=0.5\textwidth]{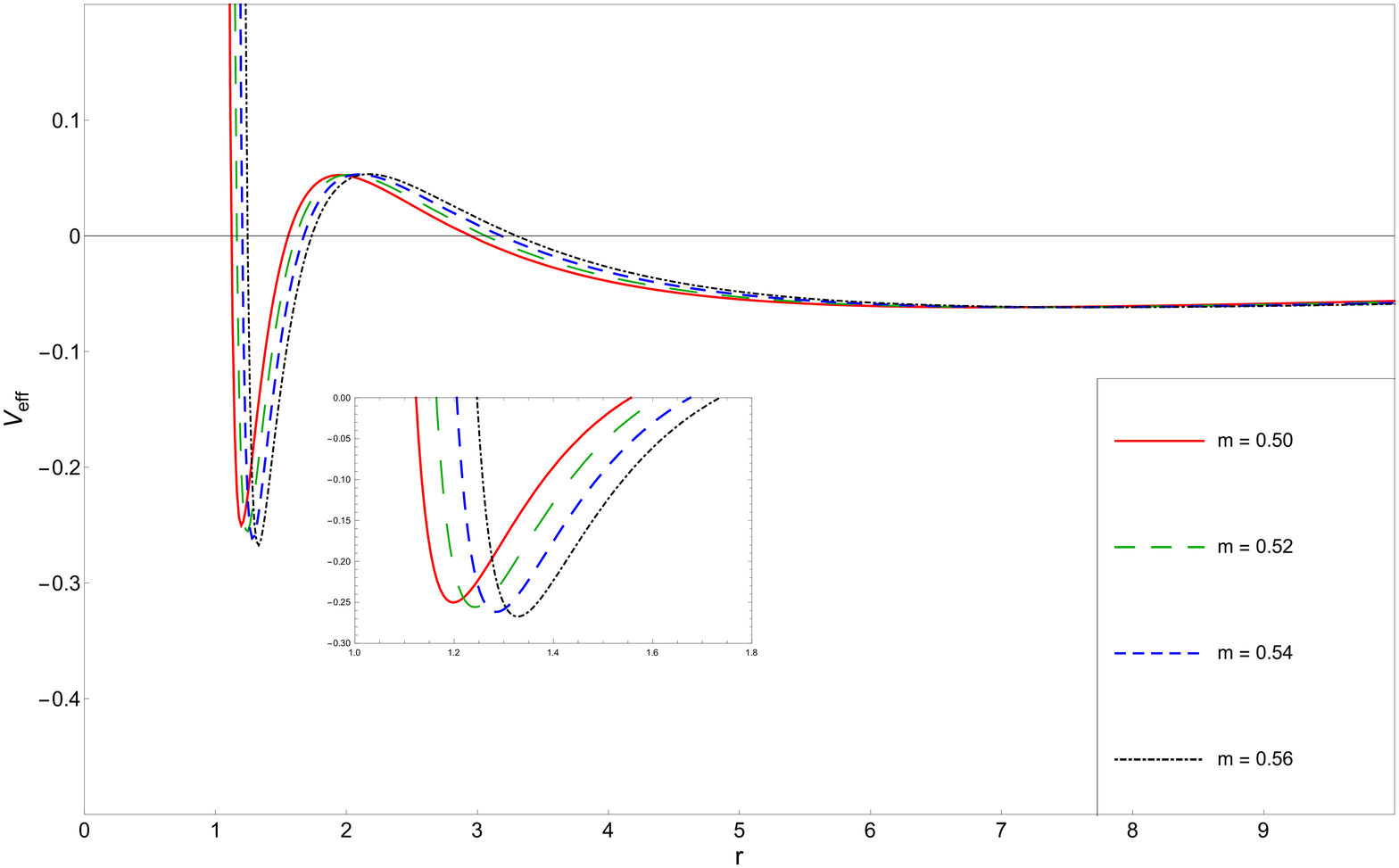}
	\\[\smallskipamount]
	\includegraphics[width=0.5\textwidth]{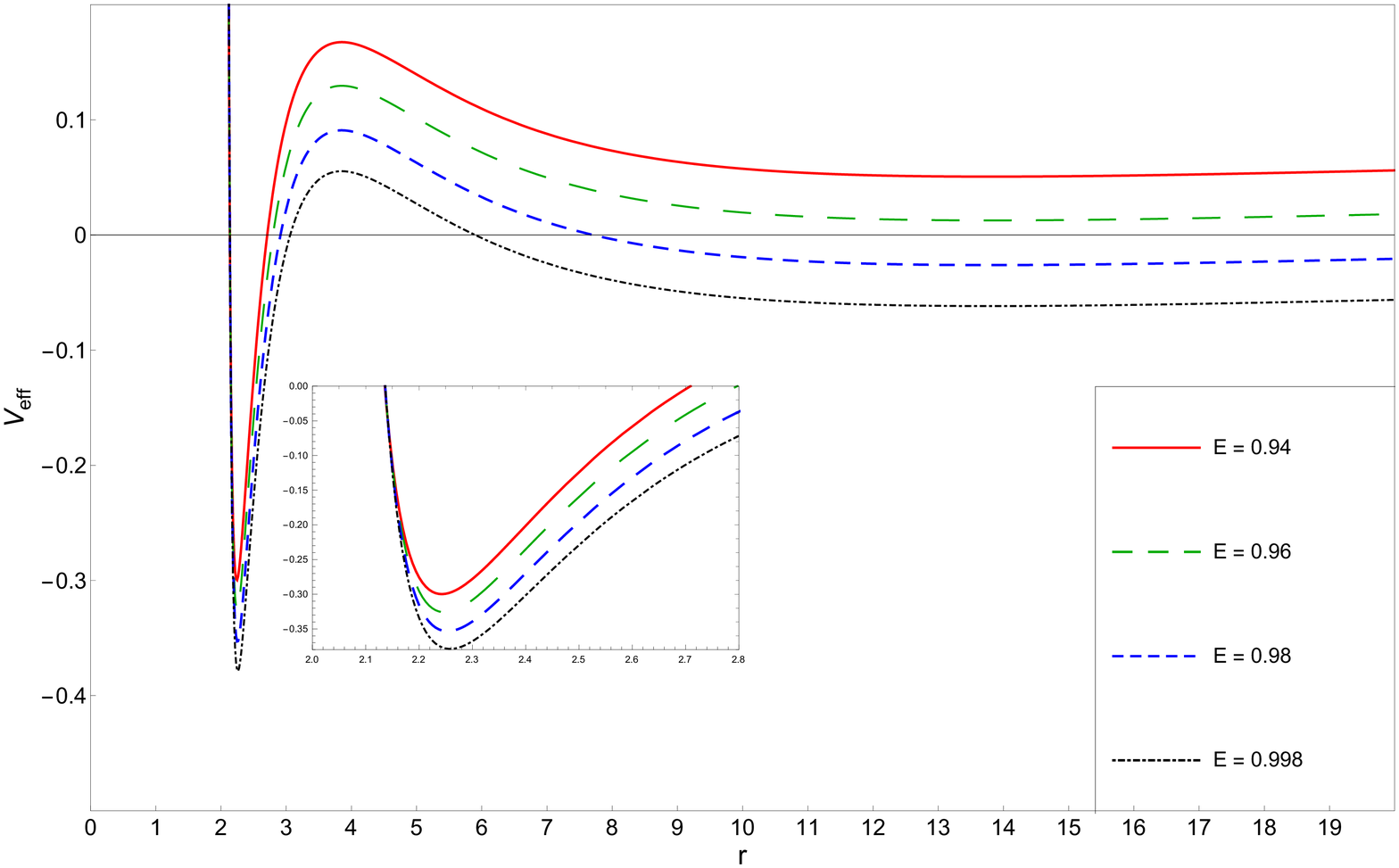}\hfill
	\includegraphics[width=0.5\textwidth]{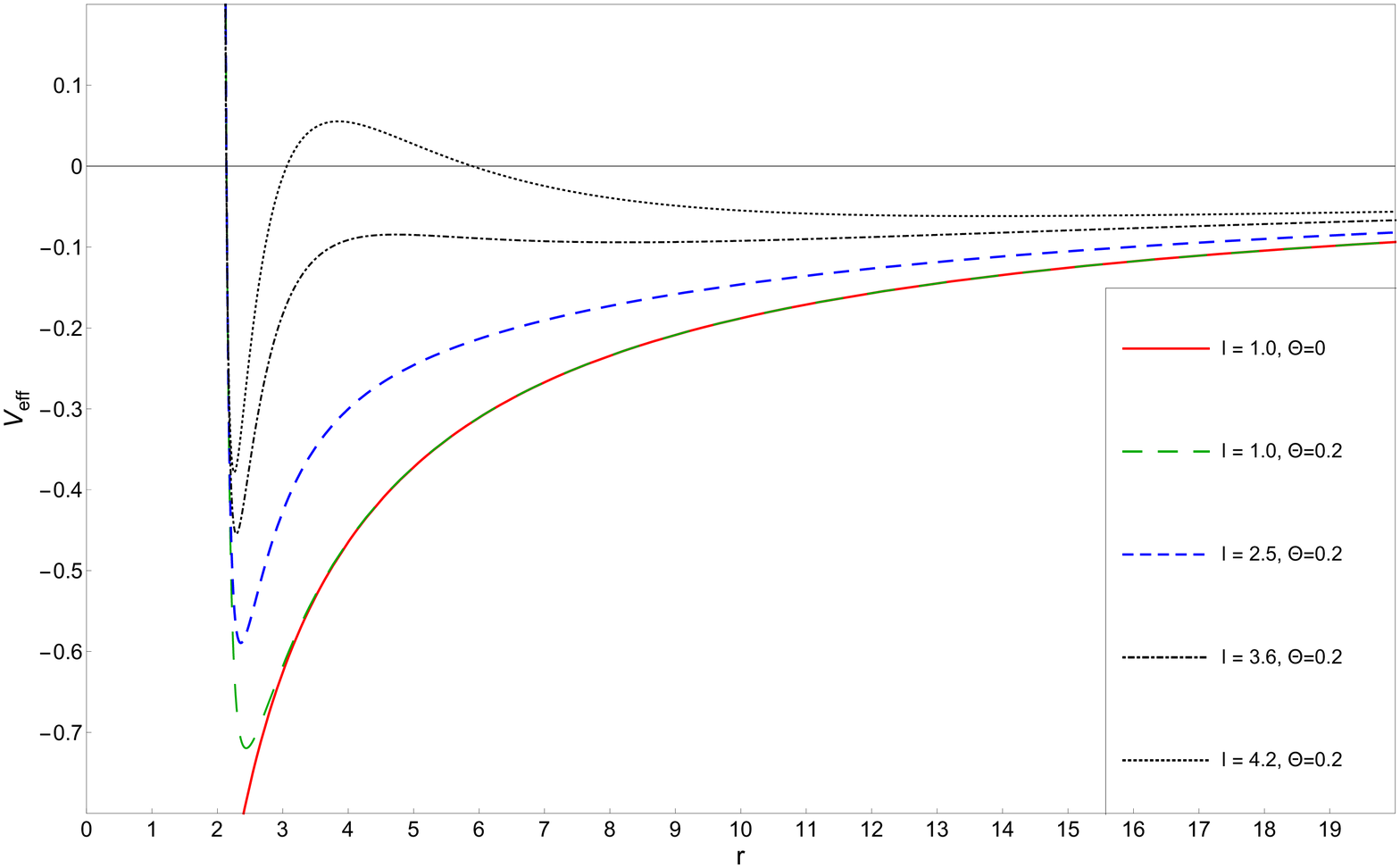}
	\end{center}
	\begin{tikzpicture}[overlay, remember picture]
	\node [] at (0.9,10) {(a)};
	\node [] at (8.4,10) {(b)};
	\node [] at (0.9,5) {(c)};
	\node [] at (8.4,5) {(d)};
\end{tikzpicture}
	\caption{The behaviors of the effective potential for massive particle. (a) different $\Theta$ and fixed: $E=0.998$, $m=1$, $l=4.2$. (b) different $m$ and fixed: $E=0.998$, $\Theta=0.2$, $l=4.2m$. (c) different $E$ and fixed: $m=1$, $\Theta=0.2$, $l=4.2$. (d) different $l$ and fixed: $E=0.998$, $\Theta=0.2$, $m=1$.  }\label{fig1}
	\end{figure}

In Figure \ref{fig1}, we show the influence of parameters ($\Theta$, $m$, $E$ and $l$) on the effective potential for a massive particle. From this figure, we can see that in the NC space-time all the extremes of the effective potential are located outside the event horizon whatever the value of the parameter used, and this deformed geometry adds a new minimum to this effective potential witch give us multiples stable circular orbits. In (a) when $\Theta$ increases the maximum peak of the curve decrease and shifted a little off the event horizon. We need to note that the divergence around the event horizon is a consequence of the non-commutative geometry, which play the role of barrier to prevent particles at the high energy from falling into the event horizon. From Fig. \ref{fig1} (b), we can see that the increase of mass shifted the effective potential off the event horizon and increases the depth of the potential well in the NC space-times. Like we see in (c) the effective potential depends on the energy of the test particle in the NC space-time \eqref{eqt3.73}, then the increase of the energy applied the decrease in the level of the effective potential and increase in the depth of the potential well. While for a particle with weak energy $E\ll1$, the new minimum of the effective potential disappear then this particle falls into the event horizon. As another new note from (d), we should mention that in the NC space-times it always exists a minimum of the effective potential near the event horizon whatever the value of the orbital momentum, and when we increase $l$ the depth of the potential well decrease and shifted to the event horizon, and the other extremes of the effective potential has been restored when we get $l_{crt}>2\sqrt{3}m$.

In this scenario the NC geometry plays the role of the potential well near the event horizon when all matter absorbed by the black-hole is compressed into this region before entering the event horizon, this leads to the formation of an accretion disk with high-density and high-temperature around this black hole and becomes very bright, this is so-called "Black Hole Accretion Disk Theory" see Ref.\cite{BHAD1,BHAD2,BHAD3,BHAD4} which is also known in astronomy as "Quasar", Ref. \cite{quasar1,quasar2}.

This new minimum showing in the behaviors of the effective potential in Fig. \ref{fig1}, can be found in other theories like a Reissner–Nordström charged black hole \cite{RN1,RN2} or in the theory of non-singularity black hole \cite{hayw2,hayw3},..etc, while this theory has a problem where this minimum is located inside the event horizon which can not be interpreted as a stable circular orbit. In our work, the non-commutativity shifted this new minimum to the outside of the event horizon, which gives us a possibility to see it as a stable circular orbit near the event horizon, we see all of that in the following section.


\subsection{Stability condition}

In which follows, we treat the circular orbits and the stability condition in the NC space-time, to see how this deformed geometry affects this class of orbits.
For that, take the case of circular orbits ($\dot{r}=0$), the corresponding effective potential must satisfy:
\begin{equation}
	V_{eff}(r,\Theta)=V^{2}(r,\Theta)-E^{2}=0\label{eqt3.15}
\end{equation}
Now, we can find the extreme of the non-commutative effective potential given by the relation \eqref{eqt3.73}, to get the stable and unstable orbit, we need to solve this equation:
\begin{equation}
	\frac{dV_{eff}}{dr}=0\label{eqt3.74}
\end{equation}
In NC space-time, the minimum value of $V_{eff}$ appears as soon as $l \geqslant 0$, (if $l=0$, in this case, the NC parameter play the role of angular momentum), this corresponds to the Newtonian case but the existence of the maximum value of  $V_{eff}$ requires a condition on the angular momentum $l$, which is given by $l_{crt}>2\sqrt{3}m$, this corresponds to the relativistic case in commutative space-time. It is shown that the gravitational field gauge theory in NC Schwarzschild geometry using Seiberg-Witten maps is equivalent to the Newtonian case and the relativistic case in commutative Schwarzschild geometry.
\begin{table}[h]
	\begin{center}
		\caption{Some numerical values of unstable circular orbit $r_{uns}$ and the multiple stable circular orbit $r_{sta}$ in the commutative and the NC case with different parameter $\Theta$ and fixed $E=0.998, l=4.2, m=1$.}\label{tab1}
		\begin{tabular}{ c c c c c c c }
			\hline
			$\Theta$	& 0 & 0.10 & 0.15 & 0.20 & 0.25 & 0.30 \\
			\hline
			$r_{sta}(internal)$	&  & 2.16349 & 2.21421 & 2.25862 & 2.29837 & 2.33435 \\
			
			$r_{uns}$	& 3.83278 & 3.83684 & 3.8419 & 3.84894 & 3.85791 & 3.86876 \\
			
			$r_{sta}(external)$	& 13.8072 & 13.8074 & 13.8076 & 13.8078 & 13.8081 & 13.8086 \\
			\hline
		\end{tabular}
	\end{center}
\end{table}

\begin{figure}[h]
	\centering
	\includegraphics[width=0.3\textwidth]{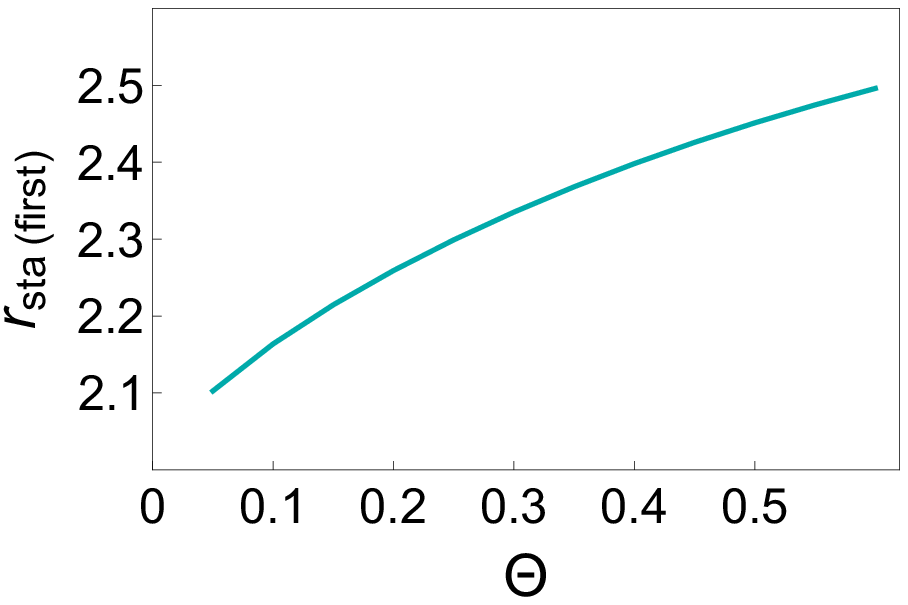}\hfill
	\includegraphics[width=0.3\textwidth]{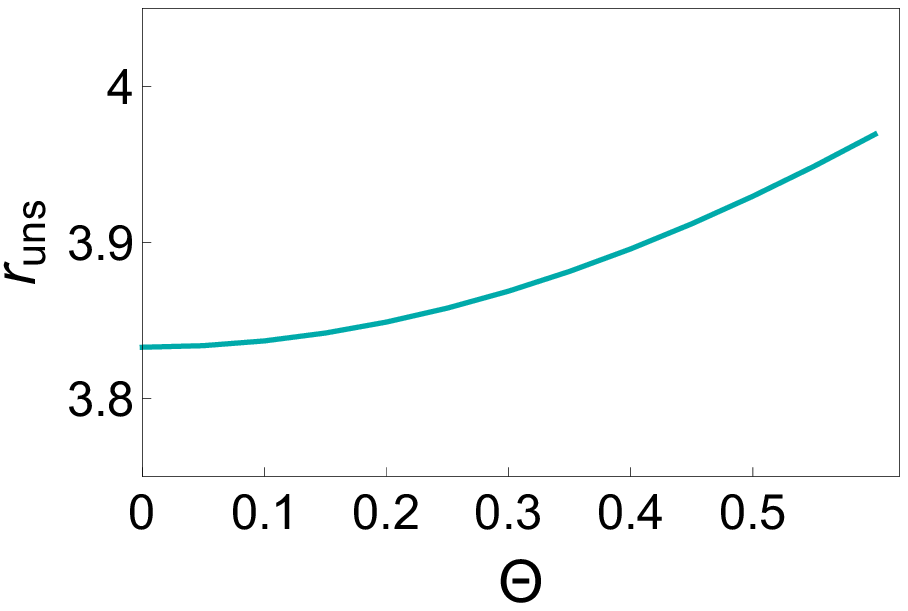}\hfill
	\includegraphics[width=0.34\textwidth]{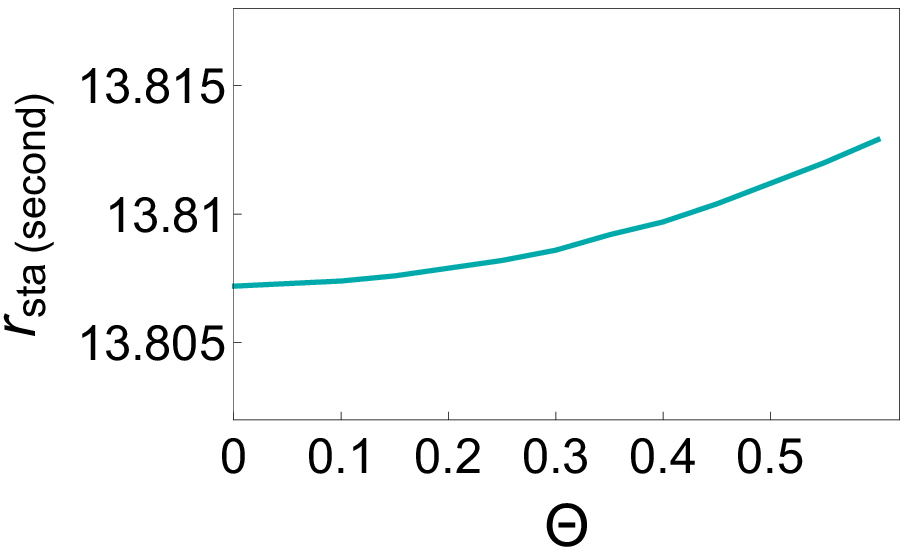}
	\\[\smallskipamount]
	\centering
	\includegraphics[width=0.3\textwidth]{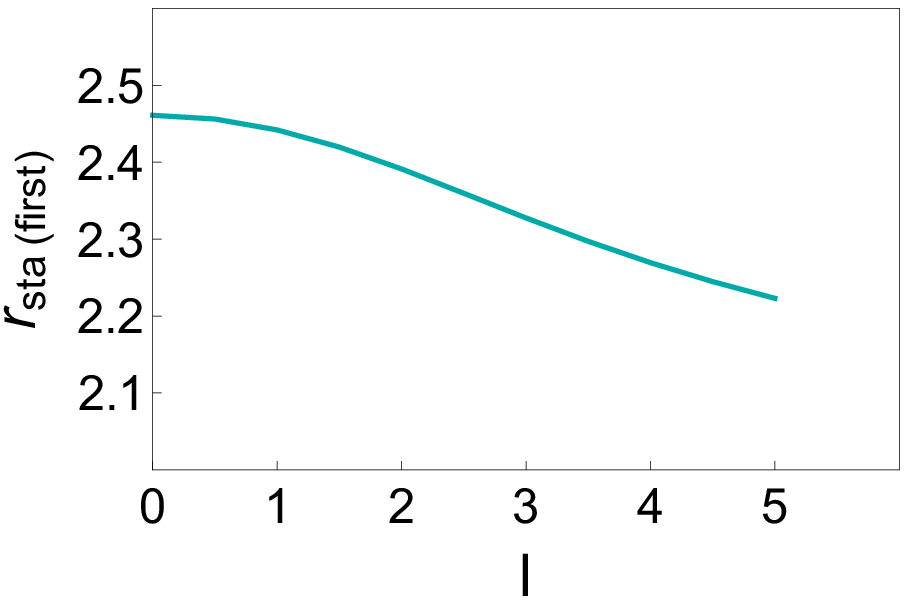}\hfill
	\includegraphics[width=0.3\textwidth]{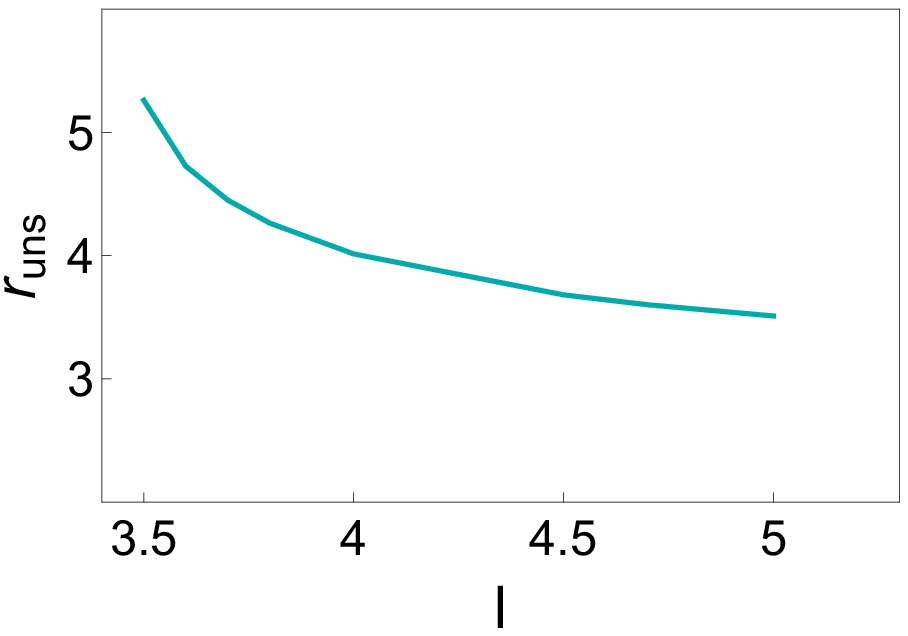}\hfill
	\includegraphics[width=0.3\textwidth]{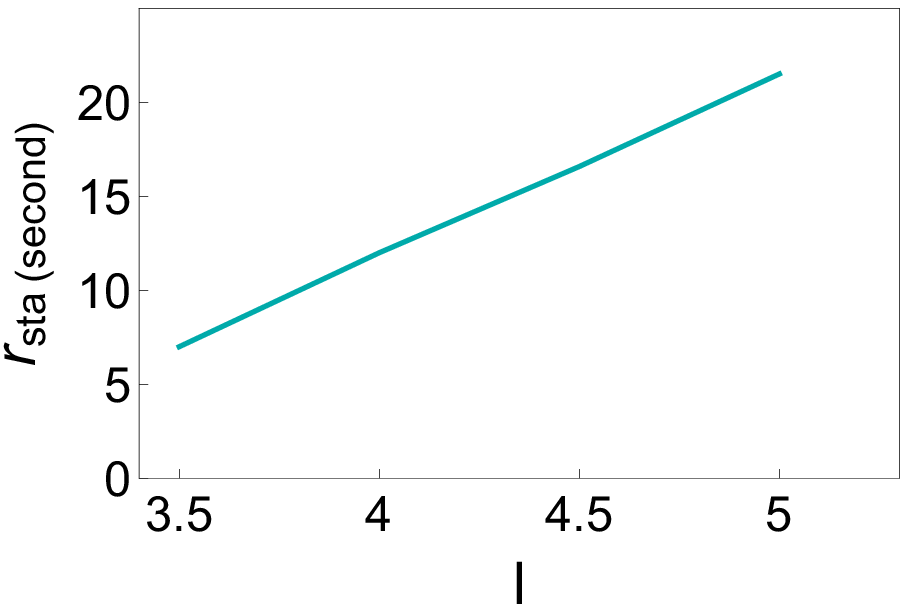}
	\begin{tikzpicture}[overlay, remember picture]
		\node [] at (-13.8,5.95) {(a)};
		\node [] at (-8.9,5.95) {(b)};
		\node [] at (-3.3,6) {(c)};
		\node [] at (-13.8,2.7) {(d)};
		\node [] at (-8.9,2.85) {(e)};
		\node [] at (-3.6,2.75) {(f)};
	\end{tikzpicture}
	\caption{The behaviors of the radius of circular orbit for a particle in the NC space-time. Unstable and multiple stable circular orbit as function of $\Theta$ and for fixed $l=4.2$, $E=0.998$, $m=1$ in (a), (b) and (c), and as function of $l$ and for fixed $\Theta=0.2$, $E=0.998$, $m=1$ in (d), (e) and (f).}	\label{fig2}
\end{figure}

This table show some numerical solution of equation \eqref{eqt3.74}, where it represents the variation of unstable and multiple stable circular orbits as a function of NC parameter $\Theta$, where the three type of circular orbit increase with the increase of $\Theta$. This behavior can be seen in Fig. \ref{fig2}.

We can conclude from Fig. \ref{fig2} that as the NC parameter $\Theta$ increases, all the types of radius increase in (a), (b), and (c). Therefore unstable circular orbital has a greater radius in NC spaces as the parameter increases, this indicates a strong gravitational field.
Where we can see as the angular momentum $l$ increase, the unstable and internal stable circular orbit decrease, while the external stable circular orbit increase in (d), (e), and (f).

In astrophysics, the innermost stable circular orbit (ISCO) has a big importance to describe the motion of a test body around a compact object. Where this class of orbits can be obtained from the stability condition, which is given by:
\begin{equation}
	\frac{d^{2}V_{eff}}{dr^{2}}\geqslant 0 \label{eqt3.75}
\end{equation}

\begin{figure}[h]
	\centering
	\includegraphics[width=0.305\textwidth]{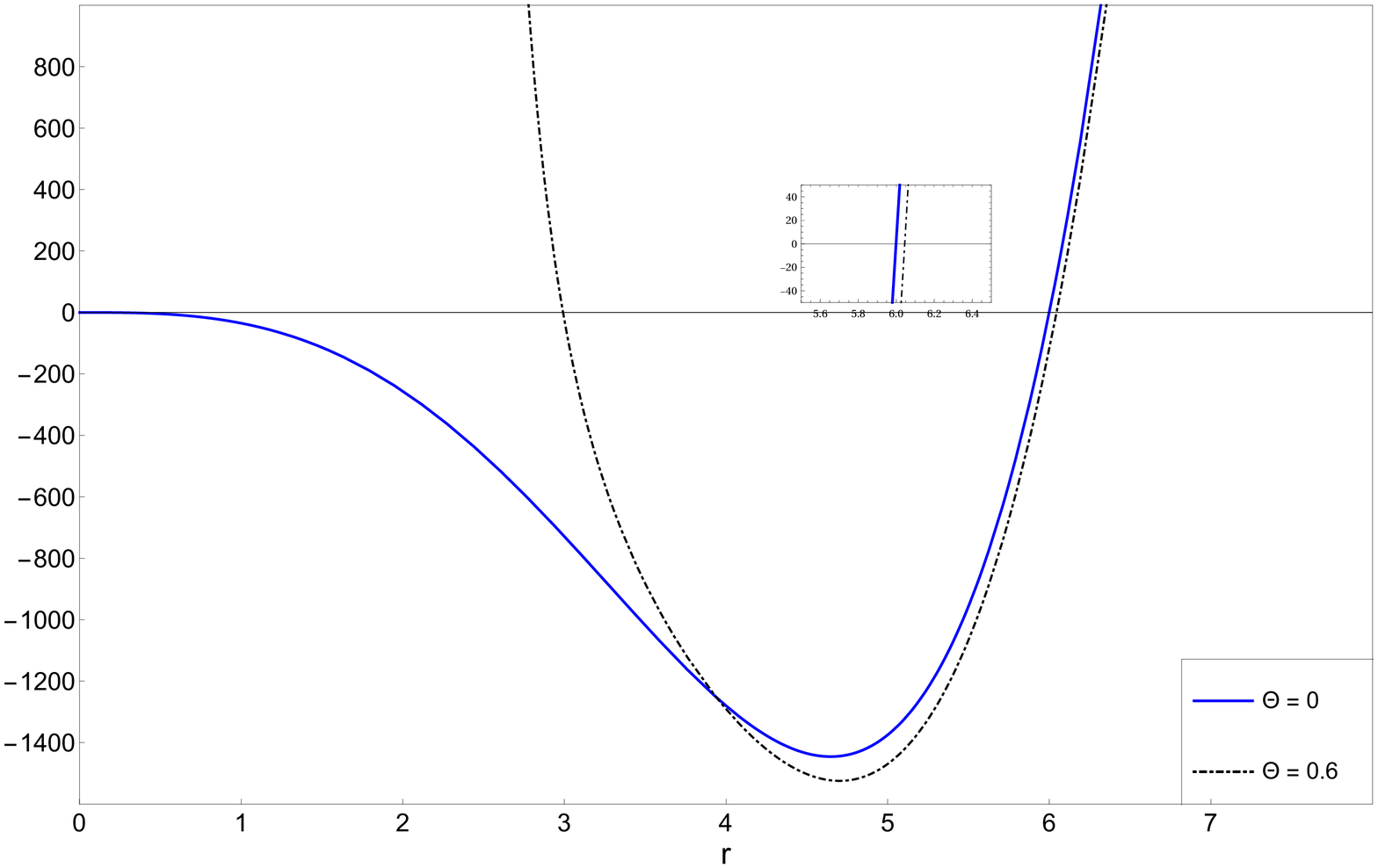}\hfill
	\includegraphics[width=0.3\textwidth]{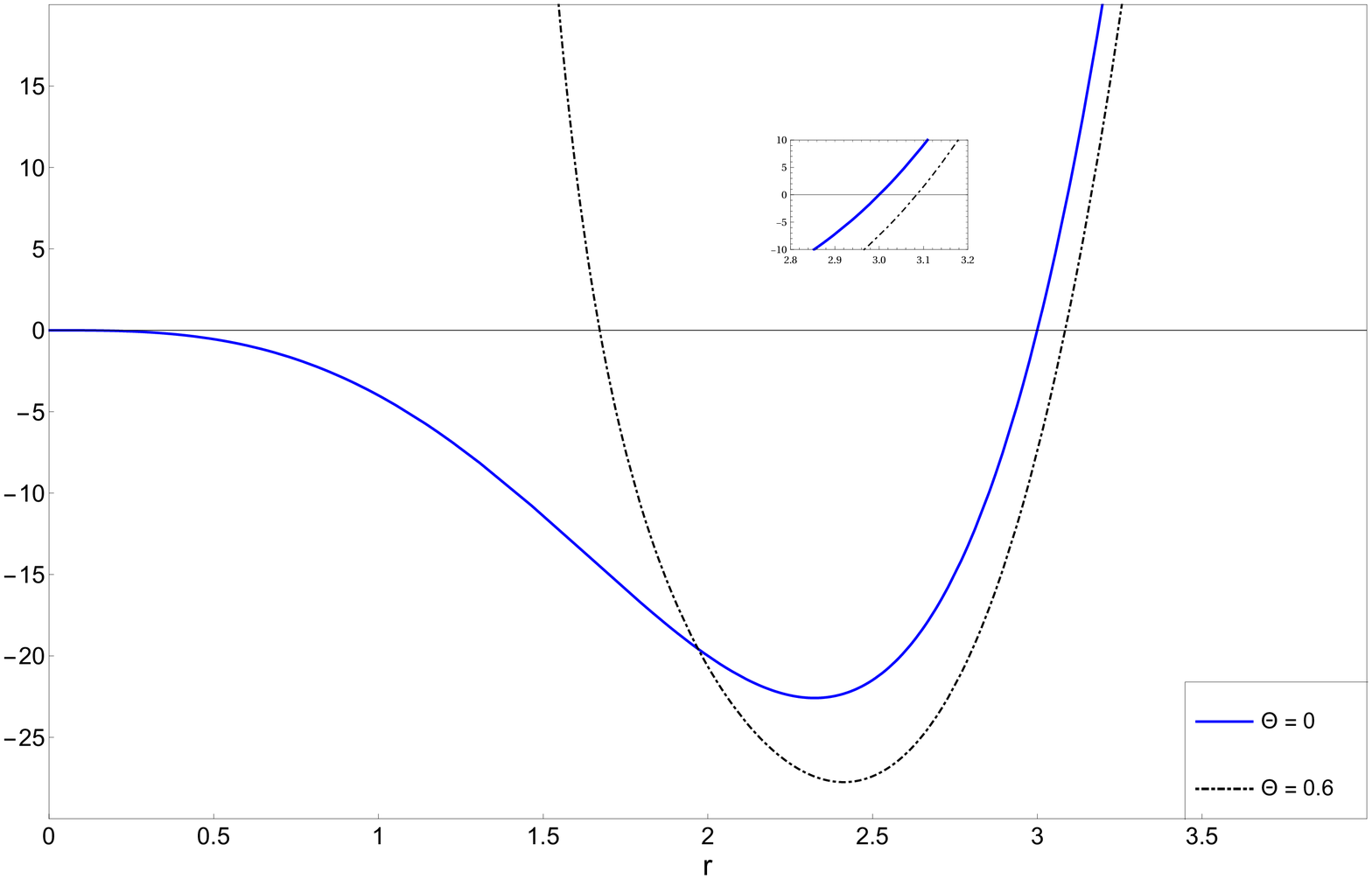}\hfill
	\includegraphics[width=0.3\textwidth]{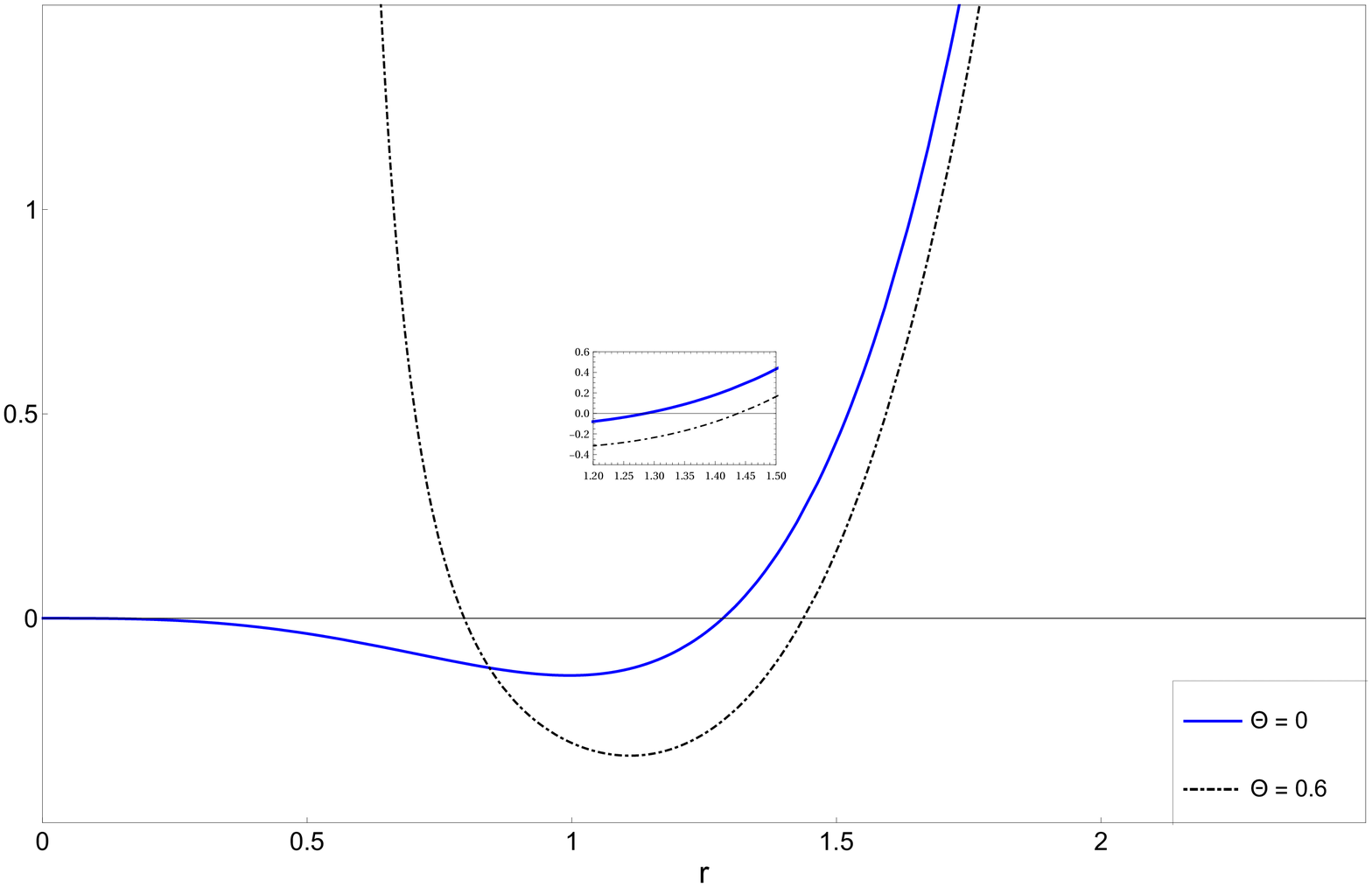}
	\begin{tikzpicture}[overlay, remember picture]
		\node [] at (-14.2,2.5) {(a)};
		\node [] at (-9.2,2.5) {(b)};
		\node [] at (-4.2,2.5) {(c)};
	\end{tikzpicture}
	
	\caption{The condition for stability of circular orbits for different $\Theta$ and fixed other parameter: (a) $E=1$, $l_{crit}=2\sqrt{3}$, $m=1$. (b) $E=1$, $l_{crit}=\sqrt{3}$, $m=0.5$. (c) $E=1$, $l_{crt}=6\sqrt{3}/14$, $m=3/14$.}
	\label{fig3}
\end{figure}

The numerical solution of these condition show that, in the commutative Schwarzschild space with $l_{crit}$, we get $r^{C}_{min}\geqslant 6$ and for a NC Schwarzschild space using the Seiberg-Witten maps withe parameter $\Theta=0.2$ we get two conditions of stability orbits (see Fig.\ref{fig3}) $r_s \ll r^{NC}_{min} \leqslant 2.46729$ and $r^{NC}_{min} \geqslant 6.00772$, which corresponding to the multiple innermost stable orbits.

The Figure \ref{fig3}, show the behaviors of the composite conditions equations \eqref{eqt3.74} and \eqref{eqt3.75} for fixed $E$ and different other parameters $l_{crit}$, $m$ and $\Theta$, as we see in the commutative space $\Theta=0$ we have just one condition for the innermost stable circular orbit, but the NC space increase this condition of the innermost stable circular orbit and he add a new condition for the stable circular orbit near the event horizon of the static black hole. Another note can be seen from the graph when we decrease the mass of the black hole we found that the NC effect increase and that suggests to the NC term correction to be in the proportionality to $(\propto\frac{1}{m})$.
\begin{figure}[h]
	\centering
	\includegraphics[width=0.5\textwidth]{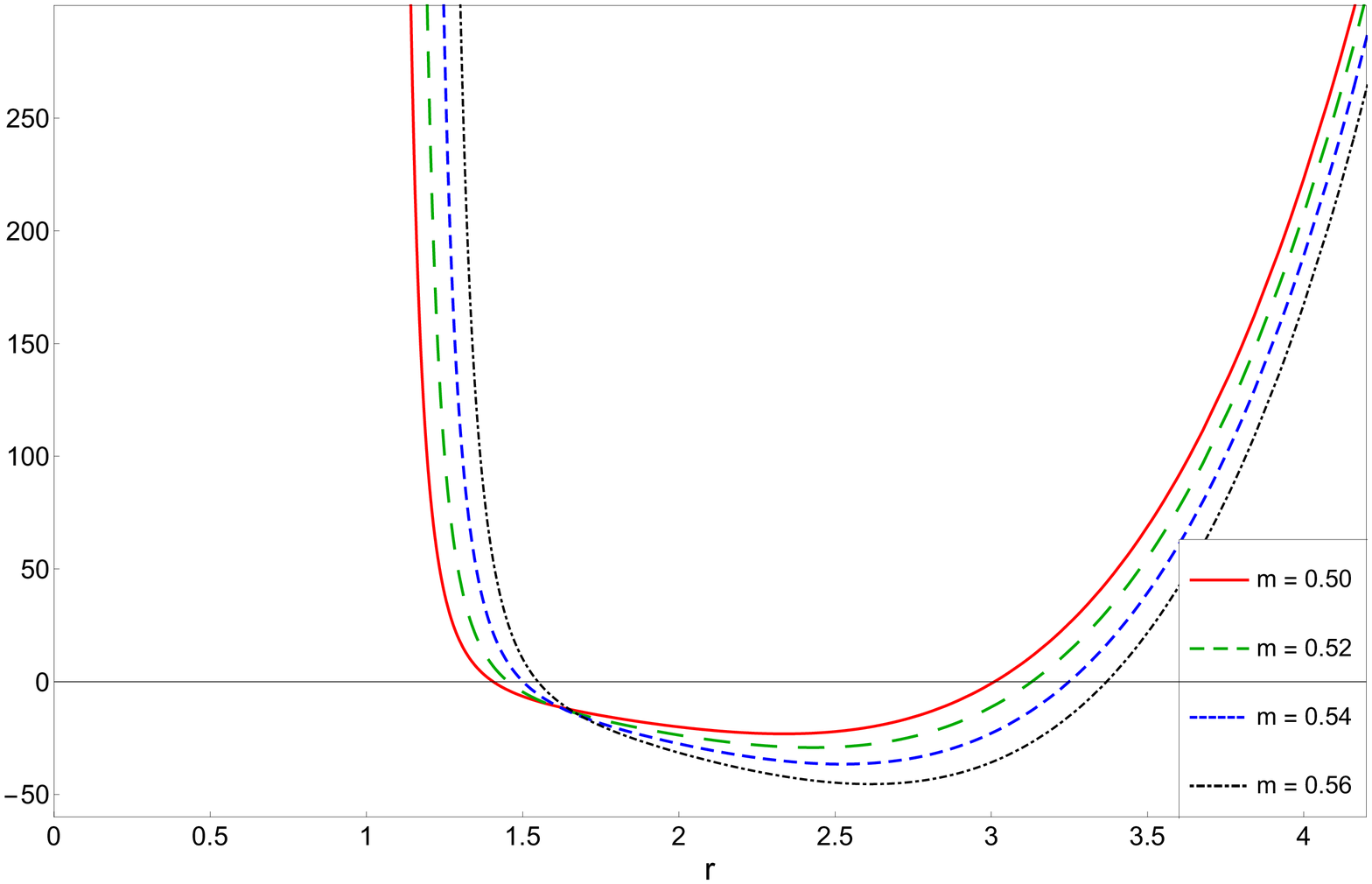}\hfill
	\includegraphics[width=0.5\textwidth]{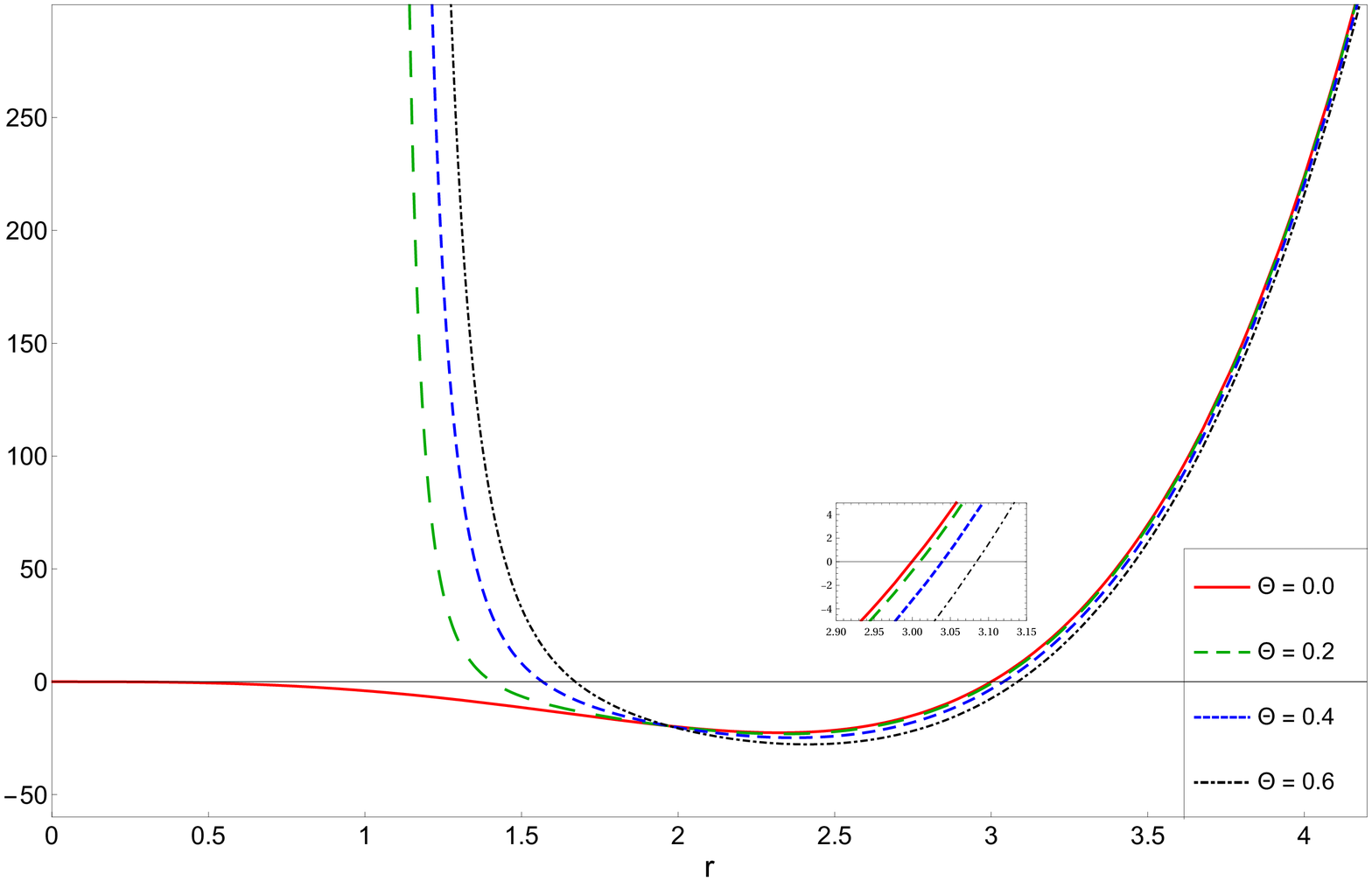}
	\begin{tikzpicture}[overlay, remember picture]
		\node [] at (-14.4,0.8) {(a)};
		\node [] at (-6.85,0.8) {(b)};
	\end{tikzpicture}
	
	\caption{The condition for the stability of circular orbits for fixed $E=1$ . (a) different $m$, $l_{crit}=2\sqrt{3} m$  and fixed $\Theta=0.2$. (b) different $\Theta$ and fixed $m=0.50$}
	\label{fig6}
\end{figure}

As we see in Fig \ref{fig6}, the behaviors for the stability condition of circular orbits as a function of the mass $m$ in (a) and as a function of the NC parameter $\Theta$ in (b) is showing, then we see with the increase of the mass the two stability condition in the NC space-times increase, and the same note when we increase the NC parameter this two stability condition increase. From this behavior in Fig. \ref{fig6}, we can see that, the NC parameter $\Theta$ plays the same role as the mass $m$, and that may be can use to explain the dark matter in this universe.

\begin{table}[h]
	\begin{center}
		\caption{Some numerical solution for the radius condition of innermost stable circular orbit with different parameter $\Theta$ and fixed $E=1,\quad l_{crit}= 2\sqrt{3}m$, $m$. (a) $m=1$, (b) $m=0.5$, (c) $m=3/14$.}\label{tab2}
		\begin{tabular}{ c c c c c c c }
			\hline
			\hline
			$\Theta$	& 0 & 0.10 & 0.15 & 0.20 & 0.25 & 0.30 \\
			\hline
			\hline
			$r_{(a)min} \geqslant$	& 6 & 6.00127 & 6.00286 & 6.00507 & 6.00792 & 6.01138  \\
			$r_s \ll r_{(a)min} \leqslant$	&  & 2.39118 & 2.48542 & 2.5655 & 2.63613 & 2.69974  \\
			\hline
			$r_{(b)min} \geqslant$	& 3 & 3.00254 & 3.00569 & 3.01008 & 3.01566 & 3.02241 \\
			$r_s \ll r_{(b)min} \leqslant$	&  & 1.28275 & 1.34987 & 1.40569 & 1.45373 & 1.49587 \\
			\hline
			$r_{(c)min} \geqslant$	& 1.28571 & 1.29157 & 1.29869 & 1.3083 & 1.32011 & 1.33377 \\
			$r_s \ll r_{(b)min} \leqslant$	&  & 0.616476 & 0.657125 & 0.688445 & 0.713273 & 0.733258  \\
			\hline
			\hline
		\end{tabular}
	\end{center}
\end{table}

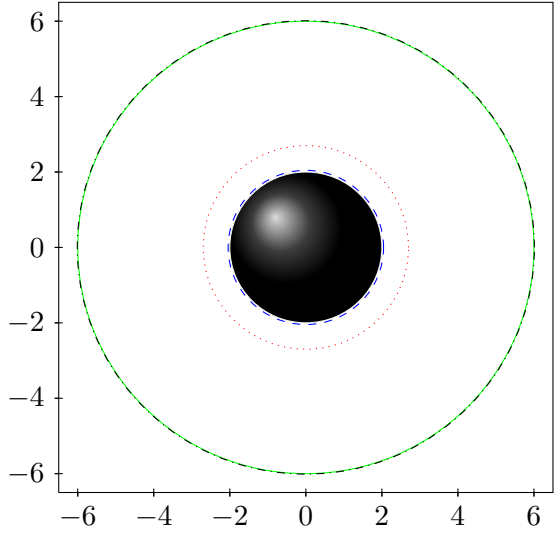
\begin{figure}[h]
	\centering
	\begin{tikzpicture}[scale=0.5]
		\draw [ color=white,ball color=black,smooth] (0,0) circle (2cm);
		\draw [ red,dotted] (0,0) circle (2.69974cm);
		\draw [green ] (0,0) circle (6cm);
		\draw [ dash dot ] (0,0) circle (6.01138cm);
		\draw [ blue, dashed] (0,0) circle (2.04406cm);
		\draw (-6.5,-6.5) -- (6.5,-6.5) -- (6.5,6.5) -- (-6.5,6.5) -- (-6.5,-6.5);
		\foreach \x in  {-6,-4,-2,0,2,4,6}
		\draw[shift={(\x,-6.5)},color=black] (0pt,3pt) -- (0pt,-3pt);
		\foreach \x in {-6,-4,-2,0,2,4,6}
		\draw[shift={(\x,-6.5)},color=black] (0pt,0pt) -- (0pt,-3pt) node[below]
		{$\x$};
		\foreach \y in  {-6,-4,-2,0,2,4,6}
		\draw[shift={(-6.5,\y)},color=black] (3pt,0pt) -- (-3pt,0pt);
		\foreach \y in {-6,-4,-2,0,2,4,6}
		\draw[shift={(-6.5,\y)},color=black] (0pt,0pt) -- (-3pt,0pt) node[left]
		{$\y$};
	\end{tikzpicture}
	\caption{The position of the innermost stable circular orbit with $E=1$, $m=1$, $h=1$, $l_{crit}= 2\sqrt{3}$, the circle with solid line represents ISCO for the Schwarzschild black hole (black disk in center) in commutative case $\Theta=0$. Dashed line represent the NC event horizon, dot lines represent the new ISCO in internal region (near the event horizon), dot-dashed lines represents ISCO in external region for the Schwarzschild black hole in the NC case $\Theta=0.3$.}
	\label{fig5}
\end{figure}

Like we see in table \ref{tab2}, some numerical solutions are obtained according to conditions equations \eqref{eqt3.74} and \eqref{eqt3.75} as show in Fig. \ref{fig3}, which represent the variation of innermost stable circular orbit radius as function of $\Theta$, which is found to increase with increasing $\Theta$. Where we see, the NC space predicted a new stable circular orbit near the event horizon, which was unknown in the commutative space, and all of this is shown in Fig \ref{fig5}.

From these two tables (\ref{tab1}, \ref{tab2}), we can conclude that the NC space increases the radius of the stable circular orbits and add a possibility of multiple stable circular orbits near the event horizon of a static black hole.

\subsection{Orbital motion}

In order to obtain the analytic formula for the periastron advance we need to obtain the equation of motion \eqref{eqt 3.68} as a function of $\phi$, for that we need to use the angular momentum equation \eqref{eqt3.60}, to write $r=r(\phi)$:
\begin{equation}
	\frac{dr}{d\tau}=\frac{dr}{d\phi}\frac{d\phi}{d\tau}=\frac{l}{\tilde{g}_{\phi\phi}(r,\Theta)}\frac{dr}{d\phi}
\end{equation}
we pot this equation into the equation \eqref{eqt 3.68}, we get:
\begin{equation}
	\left(\frac{dr}{d\phi}\right)^{2}=-\frac{\tilde{g}_{\phi\phi}^{2}(r,\Theta)}{l^{2}}V_{eff}(r,\Theta)
\end{equation}
using the relations (\ref{eqt 3.68} and \ref{eqt3.73}), with case of a massive particle $h=m_{0}^{2}$. We define a new variable $u=\frac{1}{r}$, and after some algebra we can get this finale equation:

\begin{align}
	\left(\frac{du}{d\phi}\right)^{2}&=\frac{(E^{2}-m_{0}^{2}c^{2})}{l^2}+\frac{2mm_{0}^{2}c^{2}}{l^{2}}u-u^2+2mu^3-\Theta^{2}\left\{-u^4(1-2mu)\left(\frac{5}{8}-\frac{3}{8}\sqrt{1-2mu}\right.\right.\notag\\
	&\left.\left.+\frac{1}{16}mu\left(-17+\frac{5}{\sqrt{1-2mu}}\right)+\frac{m^2u^2}{(1-2mu)^{\frac{3}{2}}}\right)-\frac{2u^2}{l^2}\left(E^2+(-1+2mu)(m_{0}^{2}c^{2}+l^2u^2)\right)\right.\notag \\
	&\left.\times\left(\frac{5}{8}-\frac{3}{8}\sqrt{1-2mu}+\frac{1}{16}mu\left(-17+\frac{5}{\sqrt{1-2mu}}\right)+\frac{m^2u^2}{(1-2mu)^{\frac{3}{2}}}\right)\right.\notag\\
	&\left.+\left(\frac{E^{2}mu^3(64u^2m^2+mu(-49+13\sqrt{1-2mu})+2(13-3\sqrt{1-2mu}))}{16l^{2}(1-2mu)^2}\right)\right.\notag\\
	&\left.+\frac{mu^3(m_{0}^{2}c^{2}+l^2u^2)(12u^2m^2+mu(-14+\sqrt{1-2mu})-(5+\sqrt{1-2mu}))}{8 l^2 (1-2mu)}\right\}+\mathcal{O}(\Theta^{4})\label{eqt3.28}
\end{align}
We use the effect that $mu\ll 1$, and we rewrite the above equation in the linear form and we stop to the 3 order in $u$, so we find:
\begin{align}
	\left(\frac{du}{d\phi}\right)^{2}&=\frac{(E^{2}-m_{0}^{2}c^{2})}{l^2}+\frac{2mm_{0}^{2}c^{2}}{l^{2}}u-u^2+2mu^3+\frac{\Theta^{2}}{2l^2}\left\{(E^2-m_{0}^{2}c^{2})u^2+m(5m_{0}^{2}c^{2}-4E^{2})u^3\right\}\label{eqt3.29}
\end{align}
Derive the above equation with respect to $\phi$, then it yields:
\begin{align}
	\frac{d^{2}u}{d\phi^{2}}+u=\frac{mm_{0}^{2}c^{2}}{l^{2}}+3mu^{2}+\frac{\Theta^{2}}{2l^2}\left\{(E^2-m_{0}^{2}c^{2})u+\frac{3m}{2}(5m_{0}^{2}c^{2}-4E^{2})u^2\right\}
\end{align}
The above equation is the non-commutative geodesic equation.

\begin{figure}[h]
	\centering
	\includegraphics[width=0.5\textwidth]{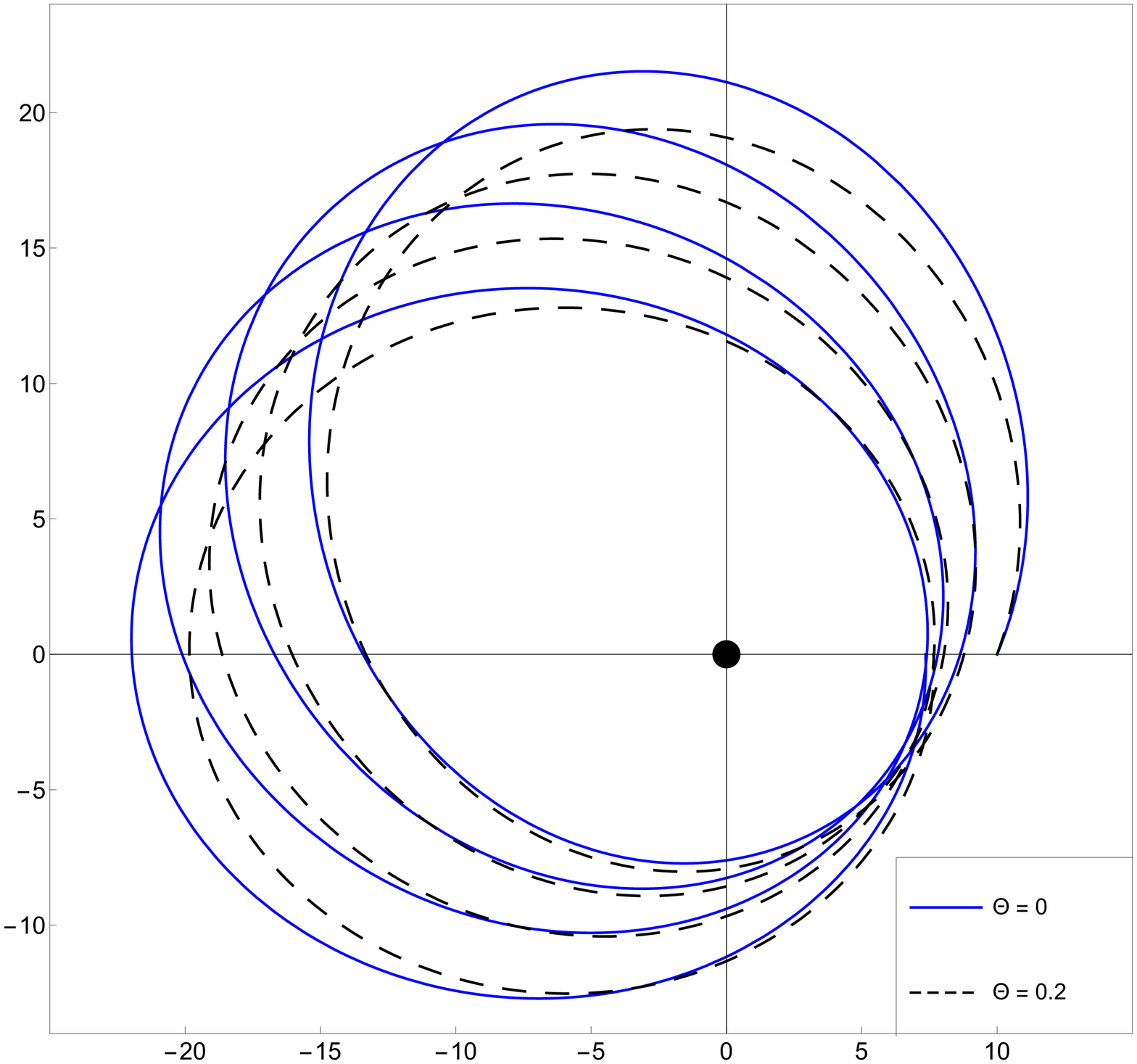}\hfill
	\begin{tikzpicture}[overlay, remember picture]
		\node [] at (-7.3,6.6) {(a)};
		\node [] at (0.8,6.6) {(b)};
	\end{tikzpicture}
	\includegraphics[width=0.45\textwidth]{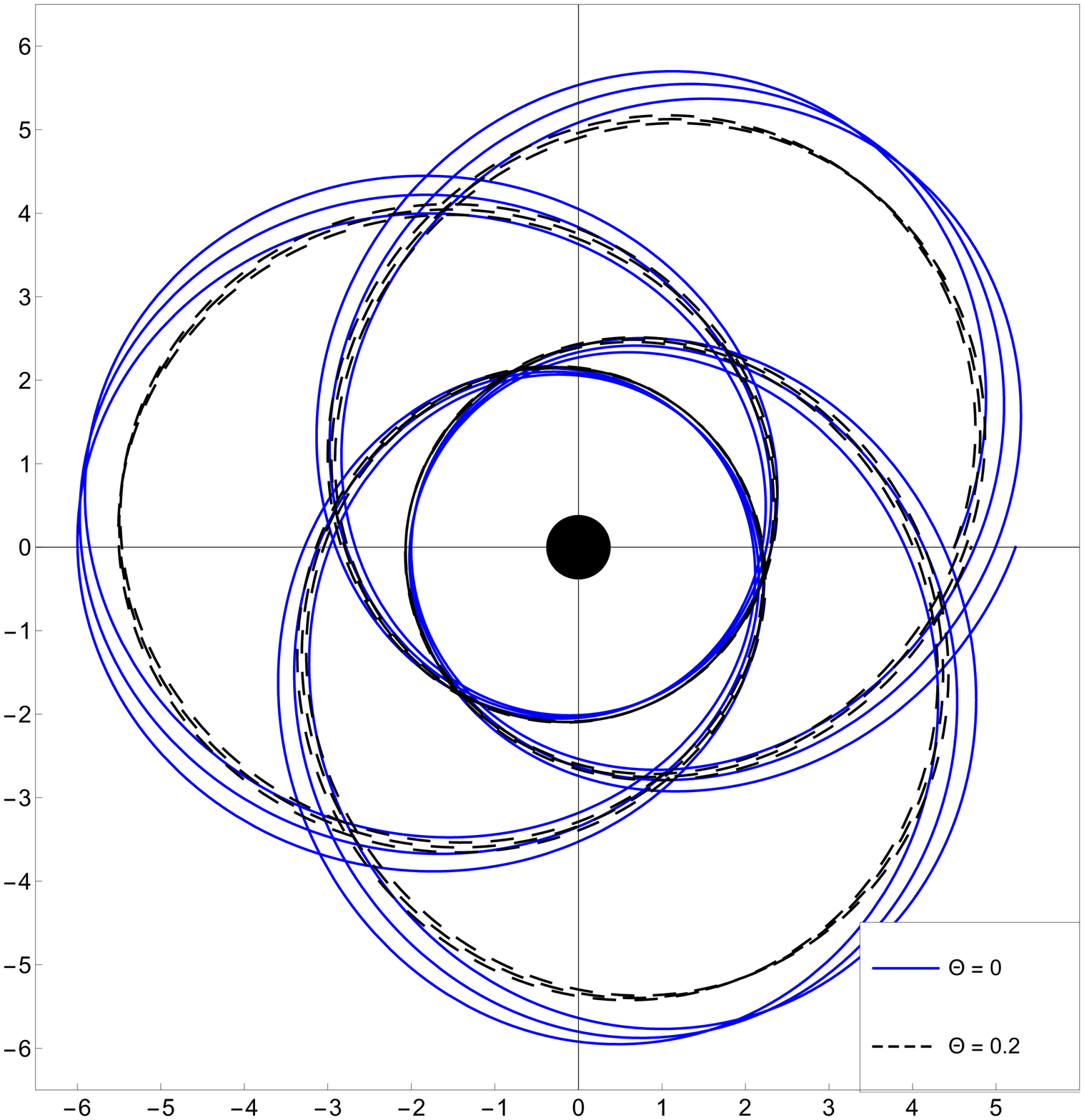}

	\caption{Time-like geodesic for a test particle $h=1$ around a non-commutative Schwarzschild Black-Hole, with different values of $\Theta$ and fixed other parameters in the plan $\theta=\frac{\pi}{2}$: (a) $M=\frac{3}{14}$, $l=1.586$, $E=0.993$. (b) $M=\frac{3}{14}$, $l=0.915$, $E=0.975$.}
	\label{fig4}
\end{figure}

In figure \ref{fig4}, we plot the geodesic equation \eqref{eqt3.29} for a massive particle around a NC Schwarzschild Black-Hole, for different values of $l$ and $E$ with a fixed black hole mass to $m=3/14$. Like we see in (a) and (b), the non-commutativity of the spacetime decreases the major semi-axis of the particle orbit, and he stays stable. This clarifies that the NC effects are responsible for strengthening the strength of the gravitational field.


\subsection{Periastron advance of Mercury orbit}
Let's write this equation into the perturbation form of the Keplerian trajectory equation:
\begin{equation}
	\frac{d^{2}u}{d\phi^{2}}+u=\frac{m}{\tilde{l}^{2}}+\frac{g(u)}{\tilde{l}^{2}}
\end{equation}
where $\tilde{l}=\frac{l}{m_{0}c}$, and $g(u)=3mu^{2}+\frac{\Theta^{2}}{2l^2}\left\{(E^2-m_{0}^{2}c^{2})u+\frac{3m}{2}(5m_{0}^{2}c^{2}-4E^{2})u^2\right\}$.
We follow the same steps in the Ref. \cite{adkins}. The angle deviation after one revolution it's given by:
\begin{equation}
	\Delta\phi=\frac{\pi g_{1}}{\tilde{l}^{2}}\label{eqt3.81}
\end{equation}
where $ g_{1}=\frac{dg(u)}{du}\mid_{u=\frac{1}{b}}$, and the distance $b$ is defined by $b=m\alpha(1-e^{2})$, with $\alpha$, $e$ denote the major semi-axis and the eccentricity of the movement. Using the relation \eqref{eqt3.81}, the angle deviation in the NC space:

\begin{equation}
	\Delta\phi=\frac{6\pi GM}{c^{2}\alpha (1-e^{2})}+\pi\Theta^{2}\left[\frac{(E^{2}_0/c^{2}-m^{2}_{0}c^{2})}{2m\alpha(1-e^{2})}+\frac{6(m^{2}_{0}c^{2}-E^{2}_0/c^{2})}{\alpha^{2}(1-e^{2})^{2}}+\frac{3m^{2}_{0}c^{2}}{2\alpha^{2}(1-e^{2})^{2}}\right]
	\label{eqt5.84}
\end{equation}
So, we found a result that is a little bit close with what was found in \cite{ulh1} where are used just the star product. But in our work, we are using the Seiberg-Witten map.
By using the relativistic relation of dispersion, we can find:
\begin{equation}
	\Delta\phi=\frac{6\pi GM}{c^{2}\alpha (1-e^{2})}+\pi\Theta^{2}\left[\frac{m^{2}_{0}v^{2}c^{2}}{2GM\alpha(1-e^{2})}-\frac{6m^{2}_{0}v^{2}}{\alpha^{2}(1-e^{2})^{2}}+\frac{3m^{2}_{0}c^{2}}{2\alpha^{2}(1-e^{2})^{2}}\right]
	\label{eqt5.85}
\end{equation}
It is clear that the first term represents the predictions of the well-known general relativity and a correction which is given according to the NC parameter.

For a numerical application, we take the case of Mercury planet. We found that the NC perihelion shift:
\begin{equation}
	\mid\delta \phi_{NC}\mid=\left(1.96689\times10^{43}\right)\Theta^{2}Kg^{2}.s^{-2}
\end{equation}
And for the General Relativity predicts and the observed perihelion shift for Mercury is given in \cite{benczik}:
\begin{align}
	\delta \phi_{obs}&=2\pi\left(7.98734\pm0.00037\right)\times10^{-8}rad/rev\\
	\delta \phi_{GR}&=2\pi\left(7.98742\right)\times10^{-8}rad/rev
\end{align}
We compered the NC correction to the observable data ($\mid\delta \phi_{NC}\mid\approx \delta \phi_{obs}$), we can estimate the value of $\Theta$:
\begin{equation}
	\Theta\approx 1.597 \times 10^{-25} s.Kg^{-1}
\end{equation}
or
\begin{equation}
	\sqrt{\hbar\Theta} \approx 1.029 \times 10^{-29} m
\end{equation}
Now, we can define a lower bound for $\Theta$:

\begin{equation}
	\mid\delta \phi_{NC}\mid\quad\leq\quad	\mid \delta \phi_{GR}-\delta \phi_{obs}\mid\approx 2\pi(1\times10^{-12}) rad/rev
\end{equation}
So we get:
\begin{equation}
	\Theta\leq 5.0553 \times10^{-28}s.kg^{-1}
\end{equation}
or
\begin{equation}
	\sqrt{\hbar\Theta}\leq 5.7876 \times10^{-31}m
\end{equation}
It's clear that the NC parameter $\Theta$ is very small and is remarkable that our result is very close to the obtained in Ref. \cite{rome1,mirza}, who used the classical mechanics in NC flat space, we need to note that our result has a difference of the order $10^{-1}$ with the result in Ref. \cite{rome1}, this difference exists because we use a curved space-time, and in the Ref. \cite{mirza} adds a new degree of freedom $\gamma$ and for the specific value of $\gamma$ he can get the same our result. This result leads us to the same conclusion as Ref \cite{rome1}, the planetary system is very sensitive to the NC parameter, in this way the NC parameter plays the role of a fundamental constant of the system to describe the microstructure of the space-time in this region. So any small change in $\Theta$ implies a sensible change to our system at a large scale.

If we compare our result with the Planck length, then we find $\sqrt{\hbar\Theta}>L_{P}$. The lower bound for the NC parameter is also have a lower bound which is the Planck scale $L_{P}$:
\begin{equation}
	\sqrt{\hbar\Theta}\leq (3.5808 \times10^{4})L_{P}
\end{equation}
Use the natural units we can obtain the upper bound of the energy:

\begin{equation}
	3.39 \times 10^{14}Gev \leq \frac{1}{\sqrt{\hbar\Theta}}
\end{equation}
and this upper bound have also a upper bound which was Planck energy $E_{P}$.


\section{Conclusions}

In this paper, we have investigated the geodesic motion of a test particle in the NC Schwarzschild space-time. By using the Seiberg-Witten map and a general form of the tetrad field for the Schwarzschild black hole we show that all the non-zero components of the deformed metric $\tilde{g}_{\mu\nu}(r,\Theta)$ acquire a singularity in the NC correction term at the value $r=2m$ where we don't see in Ref. \cite{chai1}, and this singularity in the component $\tilde{g}_{00}$ remove the singularity at the origin $r=0$ of the black hole, this result emerged naturally from the NC structure of the spacetime itself, then we get a non-singularity black hole.
And we show the event horizon in the NC is bigger than in the commutative case $r^{NC}_H>r^C_H$, so the Schwarzschild radius plays the role as the radius of the compact object inside the NC black hole.

The NC effective potential of the particles in the NC Schwarzschild spacetime is calculated and through detailed analysis new stable circular orbits appear near the event horizon. Therefore, the geodetic structure of this black hole presents new types of motion next to the event horizon within stable orbits that are not allowed by Schwarzschild spacetime. This difference around the event horizon is a result of non-commutative geometry, which acts as a barrier to prevent particles from falling into the event horizon. As we found that in NC spacetime, the commutativity parameter plays the same role as the mass of black hole, which can be used to explain dark matter .

Finally, we find that the NC space-time decreases the major semi-axis of the particles orbit, this indicates that the effects of the non-commutativity increase the strength of the gravitational field. Then we obtained the NC periastron advance of Mercury orbit and with the experimental data, we get that a $\Theta$ parameter of the order $10^{-25}s.Kg^{-1}$, gives observable deviation in the perihelion shift of Mercury, and the lower bound to $\sqrt{\hbar\Theta}$ show that the NC propriety appears before the Planck length scale. The upper bound to the energy of the order of $10^{14}Gev$ confirms that the NC properties of space-time appear at the High Energy.


\appendix

\acknowledgments

This work is supported by project B00L02UN050120190001, Univ. Batna 1, Algeria.


\bibliographystyle{unsrt}
\bibliography{biblio1}

\end{document}